\documentclass[a4paper, 11pt]{article}
\usepackage{graphicx}
\usepackage{subcaption}

\usepackage{float}		
\usepackage{amsmath}
\usepackage{amsfonts}
\usepackage{dsfont}
\usepackage{mathrsfs}
\usepackage{amssymb}
\usepackage{appendix}
\usepackage{bbm}
\usepackage{epsfig}
\usepackage[english]{babel}
\usepackage[all]{xy}
\usepackage{color}
\usepackage{cite}

\newcommand*\mcapinn[2]{\vcenter{\hbox{$\mathsurround=0pt
  \ifx\displaystyle#1\textstyle\else#1\fi\bigcap$}}}

\newcommand*\mcupinn[2]{\vcenter{\hbox{$\mathsurround=0pt
  \ifx\displaystyle#1\textstyle\else#1\fi\bigcup$}}}

\DeclareFontFamily{OT1}{pzc}{}
\DeclareFontShape{OT1}{pzc}{m}{it}{<-> s * [1.200] pzcmi7t}{}
\DeclareMathAlphabet{\mathpzc}{OT1}{pzc}{m}{it}

\usepackage[left=1.5cm,right=1.5cm,top=2cm,bottom=2cm]{geometry}
\usepackage{float}
\def\baselinestretch{1.4}
\newtheorem{theorem}{Theorem}
\newtheorem{definition}{Definition}

\newtheorem{lemma}{Lemma}

\begin{document}
\title{\bf Dynamics of Opinions with Bounded Confidence in Social Cliques: Emergence of  Fluctuations}

\author{Jiangbo Zhang\thanks{J. Zhang is  with the School of Sciences, Southwest Petroleum University, Chengdu, China and the Research School of Engineering, The Australian National University, Canberra, ACT, Australia. E-mail: jbzhang@amss.ac.cn },
Deming Yuan\thanks{D. Yuan is with the School of Automation, Nanjing University of Science and Technology, Nanjing, China. E-mail: dmyuan1012@njust.edu.cn},
Lei Wang\thanks{L. Wang is with the Australian Center for Field Robotics,  The University of Sydney, NSW 2008, Sydney. E-mail: lei.wang2@sydney.edu.au},
Claudio Altafini\thanks{C. Altafini is  with the Division of Automatic Control, Department of Electrical Engineering, Link\"{o}ping University, SE-58183 Link\"{o}ping, Sweden. E-mail: claudio.altafini@liu.se}
 and Guodong Shi\thanks{G. Shi is with the Australian Center for Field Robotics, School of Aerospace, Mechanical and Mechatronic Engineering, The University of Sydney, NSW 2008, Sydney. E-mail: guodong.shi@sydney.edu.au}		
}
\date{}
\maketitle

\begin{abstract}
In this paper, we  study the evolution of opinions over social networks with    bounded confidence in social cliques. Node initial opinions are independently and identically distributed; at each time step, nodes review the average opinions of a randomly selected local clique. The  clique averages may represent local group pressures on peers. Then nodes update their opinions under bounded confidence: only when the difference between an agent individual opinion and the corresponding local clique pressure is below a threshold,  this agent opinion is updated according to the DeGroot rule as a weighted average of the two values. As a result, this opinion dynamics is a generalization of the classical Deffuant-Weisbuch model in which only pairwise interactions take place.   First of all,  we prove conditions under which all node opinions converge to finite limits. We  show that in the limits the event that all nodes achieve a consensus,  and the event that all nodes achieve pairwise distinct limits, i.e., social disagreements,  are both nontrivial events. Next, we show that opinion fluctuations may take place   in the sense that at least one agent  in the network fails to hold a converging opinion trajectory. In fact,  we prove that this fluctuation event happens with a strictly positive probability, and also constructively present an initial value event   under which the fluctuation event arises with probability one. These results add to the understanding of the role of bounded confidence  in social opinion dynamics, and the possibility of fluctuation reveals  that bringing in cliques in Deffuant-Weisbuch models have fundamentally changed the behavior of such opinion dynamical processes.
\end{abstract}



\section{Introduction}
Today in our society, social interactions among peers increasingly  take place over online social networks.  Such interactions are much broad than classical social interactions among family members, friends, co-workers, etc. Peers from various places of the world via online platforms  gather in places such as  Facebook interest groups, Twitter/Reddit discussion threads, etc., to exchange opinions about various social, economical, or political issues \cite{lat96,social-media}. The study of the underlying  dynamics of the opinion flows is of growing importance \cite{Ace2,tempo1,tempo2}, for which the classical  DeGroot model  shed lights on understanding the mechanism behind trustful social interactions and how a connected and trustful social structure  leads to a consensus or aggrement among the social members
\cite{Degroot_1974}.

In standard DeGroot model, peers  hold  opinions   described as real-valued dynamical  states, communicate with neighbors in a fixed graph representing the social network structure, and update the states iteratively at discrete slots by averaging the neighbor states that are being communicated \cite{Degroot_1974}. It was proven that as long as the underlying social graph is connected, all peer states converge to a common value
known as the consensus state. Generalizations of this DeGroot model to continuous-time dynamics and   time-varying network structures  have been extensively studied in the literature,  e.g., \cite{jad03,saber04,fax04,moreau05,Lo1,cao08,nedic09}. Since DeGroot rule imposes non-expansiveness  of the convex hull of the node opinions over time, such convergence to consensus has been proven to be true for a number of  deterministically  switching networks e.g., \cite{jad03, moreau05,cao08}.  The varying network structure can also be modeled as random graph processes, over which DeGroot types of opinion dynamics were shown to continue to lead to consensus in the mean-square or almost sure sense e.g., \cite{Fagnani,BTouri,Shi-TIT}. For both deterministic and random switching networks,  a minimum degree of connectivity is required despite the social graph may never be connected for any given time.

The strong consensus preserving property of DeGroot model under connectivity   made it extremely useful in explaining collaborative interpersonal relations and  the resulted social learning  \cite{golub,Ace2}. However, consensus is rarely observed in real-world human groups even though the underlying social network is well connected. Beyond consensus, social opinion formations may be {\it clustering} in the sense that   agent opinions converge to  distinct finite limits; {\it fluctuating} in the sense that agent opinions experience lower and higher values for all time instead of converging. In the literature, there have been quite a few proposals that started from social phenomena such as antagonism/mistrust, stubbornness, biases, etc., and went on to establish asymptotic opinion formations that are beyond consensus. In \cite{Altafini_TAC13,Shi-SIREV,Shi-OR,bullo20}, signed networks were used to model social networks with both trustful and mistrustful links, and clustering into bipartite groups was established for  DrGroot model with negative links  under structurally balanced graphs. In \cite{Ace1,Ace3}, a type of randomized  DrGroot model was studied  in the presence of stubborn agents who never revise their opinions, and it was shown that agent opinions undergo fluctuations between the stubborn opinions. It turned out fluctuations may also be observed for opinion dynamics over signed networks \cite{Shi-OR,Shi-SIREV,bullo20}, where the interplay between positive and negative links may yield such opinion formations.
 In \cite{dandekar,bias}, individual biases  were modeled as nonlinear wights on self-opinion and local group opinion in the iterations, based on which clustering to extreme opinions was also revealed.

Along this line of research, there has been an important development on bounded confidence in social interactions.  Bounded confidence attempts to capture the social tendency that peers are more inclined to
only {\it believe}  others whose opinions are within a vicinity of their own opinions, despite being {\it exposed} to  opinions in diverse ranges. There have been mainly two types of bounded confidence models.
In the Deffuant-Weisbuch model \cite{Deffuant-2000}, peers  meet randomly in pairs and exchange their opinions, but only revise their opinions by the DeGroot rule when their opinion difference is lower than a threshold. In the Hegselmann-Krause model \cite{HK02Opi,Blon10,HK19}, each  agent averages the opinions of peers whose opinions  differ below a threshold as their new belief.  The bounded confidence Deffuant-Weisbuch and Hegselmann-Krause models preserve this non-expansive property of the network states, and thus convergence of individual states is expected \cite{Deffuant-2000,HK02Opi,Blon10,HK19}. The presence of bounded confidence, however, may forbid the node states from converging to a consensus in general. There has been a large number of literature, where thorough numerical studies and excellent analytical results establish that social opinions in Deffuant-Weisbuch and Hegselmann-Krause models often converge to clusters \cite{Frasca12,LTH19,YH14,HK19,Lo2,Lo3,como2011,Hendrickx-TAC-Krause,Noor20,Bullo12,Blon10,HK02Opi,Deffuant-2000}.

 In this paper, we propose and study  opinion dynamics over  a social network   with    bounded confidence in social cliques. Social cliques are local complete subgraphs in a social network. At each time step, nodes compute  the average opinions of a randomly selected clique with a given cardinality, as a representation of peer pressure in local social networks.  Then  nodes update their opinions by averaging their current opinion and the cliqure peer pressure,   when the difference between the two is below a prescribed bound. The initial node opinions are randomly assigned, and this clique bounded confidence model is a generalization to  the classical  Deffuant-Weisbuch   model where only pairwise interactions were allowed. First of all,  we present and prove conditions under which all node states converge to  finite limits, and show consensus and disagreement clustering are both nontrivial events. Next, we prove  that fluctuations may take place for the node opinions in the sense that at least one node state in the network fails to converge to a limit value. In particular, we show that this fluctuation event happens with a strictly positive probability, and also constructively present an initial value event for the network initial opinions, under which fluctuation arises with probability one. These results add to the understanding of bounded confidence models in social opinion dynamics, and the possibility of fluctuation reveals  a new type of social opinion formations that is arguably better matching our real-world experiences.

The remainder of the paper is organized as follows. In Section \ref{sec2}, we present the social network model for our study, and introduce our problems of interest. Then Section \ref{sec3} presents our main results. Finally Section \ref{sec5} concludes the paper with a few remarks on potential future directions. All proofs of our statements are given in Appendix.

\section{Problem Definition}\label{sec2}

In this section, we propose a social network model for  bounded confidence   in social cliques, where peers in a social network randomly  interact with each other in  {\it  cliques}, i.e., local complete subgraphs \cite{clique0}; and then define our problems of interest.

\subsection{The   Social Network Model}

Consider a social network of $n$ nodes (peers) indexed in the set $\mathrm{V}=\{1,2,\dots, n\}$. Time is slotted at $t=0,1,2,\dots$. At each time $t$, each node $i\in\mathrm{V}$ randomly selects $m$ $(1\leq m\leq n)$   nodes as its neighbor from the network node set $\mathrm{V}$, independent with other nodes' selections. This results in a random set of neighbors, termed a social {\it clique} and denoted by $\mathcal{N}_{i}(t)$,  for $i\in\mathrm{V}$ and $t=0,1,\dots$. Let $\mathfrak{N}=\{\mathrm{V}_{1},\dots,\mathrm{V}_{z}\}$ be the set containing all subsets of $\mathrm{V}$ with $m$ elements, where $z=$${n}\choose{m}$ with ${n}\choose{m}$ representing the m-combinations of $\mathrm{V}$. For the random neighbor set $\mathcal{N}_{i}(t)$, we impose the following assumption.

\medskip

\noindent {\bf A1}.
The $\mathcal{N}_{i}(t)$ are independent and identically distributed for $t=0,1,\dots$, and  $
\mathbb{P}\big\{\mathcal{N}_{i}(t)=\mathrm{V}_{k}\big\}>0
$ for all $i=1,\dots,n$ and $k=1,\dots,z$ at any given time $t$.

\medskip

Each node $i$ holds an opinion $x_{i}(t)\in \mathbb{R}$ at time $t$. After interacting with the neighbors in the set $\mathcal{N}_i(t)$, each node $i$   observes the following clique  opinion as the average of the peers' opinion in the group:
\begin{equation*}
\mathbf{y}_{i}(t)=\frac{1}{m}\sum_{j\in\mathcal{N}_{i}(t)}x_{j}(t).
\end{equation*}
Then the nodes update their opinions for time $t+1$  according to
\begin{align}\label{mode}
x_{i}(t+1)=\left\{
\begin{array}{ll}
(1-\delta)x_{i}(t)+\delta\mathbf{y}_{i}(t), & \textrm{ if } |x_{i}(t)-\mathbf{y}_{i}(t)|\leq\eta;\\
x_{i}(t), & \textrm{otherwise}
\end{array}
\right.
\end{align}
for all $i\in\mathrm{V}$. Here $0<\delta<1$ is the mixing parameter and $\eta>0$ is the confidence level, which  are assumed to be two constants. For the initial node opinions $x_{1}(0),\dots,x_{n}(0)$, we impose the following assumption.

\medskip

\noindent{\bf A2}. The $x_{i}(0),i\in\mathrm{V}$ are independent and identically distributed in $[0,1]$ by a uniform distribution.

\medskip

The assumptions {\bf A1} -- {\bf A2} are assumed throughout the paper as our standing assumptions, without specific futher mentioning.

\subsection{Related Work}
The proposed social network model with clique  bounded confidence is a generalization of the  classical  Deffuant-Weisbuch type of social interactions \cite{Deffuant-2000}. In Deffuant-Weisbuch models \cite{Deffuant-2000}, peers only meet randomly in pairs, while $\mathcal{N}_{i}(t)$ is assumed to be a clique  with $m$ nodes. The boundedness of social confidence in Deffuant-Weisbuch models is inherited in our model, where the clique  opinions describe peer pressure in a local social group. When $m$ is reduced to two, our  model   recovers the Deffuant-Weisbuch model with homogeneous confidence bound \cite{Deffuant-2000}.  Another closely related  bounded-confidence model is the Hegselmann-Krause model \cite{HK02Opi,Blon10,HK19}, where at each round,  nodes average their states among a deterministic neighborhood determined by the nodes sharing the states within a given bound.  The confidence bound thus leads to  a state-dependent   communication graph, in contrast to static or time-dependent communication graphs \cite{Degroot_1974}.

The random clique  selection process is also a generalization of the gossip processes where node interactions are held between pairs  \cite{Kempe,Boyd-TIT-2006}. The advantage of utilizing cliques in a gossip process to accelerate information dissemination or computation has been noted in \cite{clique1,clique2,clique3}.

\subsection{Problems of Interest}

We are interested in the asymptotic behaviors of the node opinions from a probabilistic point of view. We use $\mathbb{P}$ to denote the probability of the total randomness generated by both the neighbor selection process and the nodes' initial values. The following example illustrates that the proposed   bounded confidence model in social cliques may undergo drastically different behaviors compared to the typical  Deffuant-Weisbuch and Hegselmann-Krause models, in the sense that node states may fail to converge for certain range of parameters.

\medskip

\noindent{\bf Example 1}. Consider a network of $n=20$ nodes. Let $m=4$ be the size of  cliques  for social interactions. Take $\delta=0.5$. Two samples paths for the node opinions $x_i(t)$ are shown in Fig. 1, respectively, for $\eta=0.3$ and $\eta=0.2$.

Clearly, when $\eta$ went from $0.3$ to $0.2$, the network opinions witnessed a phase transition from convergence to a global  consensus to random fluctuations. \hfill$\square$

\medskip

In view of  Example 1, we are interested in the following questions on the state evolution of the social dynamic model (\ref{mode}):
\begin{itemize}
    \item [{\bf Q1}.] Are there conditions on the network parameters $(m,n,\delta,\eta)$ so that all node states converge to finite limits with probability one?

    \item [{\bf Q2}.] What are the probabilities of the limiting values agreeing or disagreeing when convergence is guaranteed?
    \item[{\bf Q3}.] Can we establish ranges on the network parameters under which opinion  fluctuations emerge from the dynamical process almost surely?
\end{itemize}
Answers to these questions will add to understandings of social opinion dynamics with bounded confidence. In particular, the  almost sure fluctuations for node opinions were only observed or proved  in the literature for opinion dynamics over a type of signed social networks \cite{Shi-OR,Shi-SIREV,Altafini_TAC13}. The proposed model therefore might potentially shed lights on the study of social interaction  mechanisms  leading to non-convergent opinion formations, as in the real world convergence and consensus are rarely observed for public opinions.

\begin{figure}[H]
  \centering
  \includegraphics[scale=0.4]{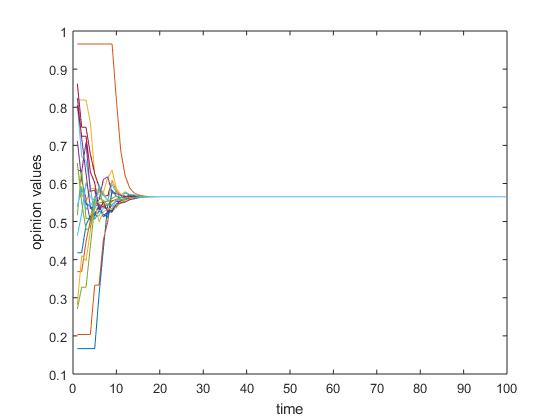}
  \label{Fig.1(a)}
  \includegraphics[scale=0.4]{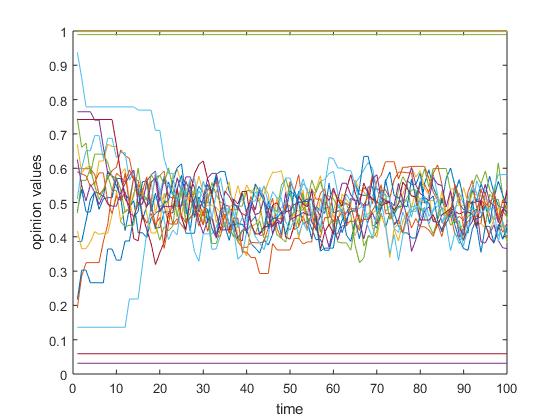}
  \label{Fig.1(b)}
  \caption{Typical sample paths of node opinions  evolution with $\eta=0.3$ (left) and $\eta=0.2$ (right) with randomly selected initial values.}
\end{figure}

\section{Main Results}\label{sec3}
In this section, we present the  results  on the asymptotic behavior of our bounded confidence opinion dynamics model (\ref{mode}).
\subsection{Almost Sure Convergence Conditions}
First of all, we present the following result on the conditions under which the network node opinions all converge to finite limits.

\begin{theorem} \label{allselected-converge}
Suppose $m=n$. Then the opinion dynamics (\ref{mode}) leads the node states to a convergence almost surely. To be precise, there exist random variables $B_{1},\dots,B_{n}$ such that
\begin{align*}
\mathbb{P}\Big\{\lim_{t\to\infty}x_{i}(t)=B_{i},i\in\mathrm{V}\Big\}=1.
\end{align*}
\end{theorem}

Theorem \ref{allselected-converge} shows that if the sampling of  cliques  is always across the entire network, all node states will converge to finite limits. This result is consistent with the studies on the Deffuant-Weisbuch and Hegselmann-Krause models.  Further, we have the following result showing that the probabilities  of having  consensus or pairwise distinct limiting states are nontrivial.

\begin{definition}
The events $\mathsf{E}_{\rm consensus}:=\{B_{1}=B_{2}=\dots=B_{n}\}$ and $$
\mathsf{E}_{\rm disagreement}:=\{B_{1},B_{2},\dots,B_{n}\ \mbox{are pairwise distinct}\}$$ are termed,
respectively, the consensus event and the disagreement clustering event  along the opinion dynamics (\ref{mode}).
\end{definition}

\begin{theorem}\label{thm2}
Suppose $m=n$ and $n\geq 4$. Let $\eta<{1}/{(n+1)}$.    Then both $\mathsf{E}_{\rm consensus}$ and $\mathsf{E}_{\rm disagreement}$ are nontrivial events, i.e., they take place with a strictly positive probability.
 \end{theorem}

Theorem \ref{allselected-converge} and Theorem \ref{thm2} are certainly quite restrictive since they only apply to the case with $n=m$. This condition $n=m$ allows us to thoroughly develop an approach to fully decompose the event space of $x_i(t),i\in\mathrm{V}$ according to the initial values, under which the kinds of results in the two theorems only become possible.  Next, we introduce the following definition on the ordered statistics of the node opinions.
\begin{definition}\label{definition-Order-statistics}

(i) The {\it ordered statistics} of opinion states $x_{i}(t), i\in\mathrm{V}$ is defined as
$\mathbf{x}_{[1]}(t),\mathbf{x}_{[2]}(t),\dots,\mathbf{x}_{[n]}(t)$,  where  $\mathbf{x}_{[k]}(t)$ represents the $k$'th smallest  values among the $x_{i}(t), i\in\mathrm{V}$, i.e.,  $\mathbf{x}_{[1]}(t)\leq\mathbf{x}_{[2]}(t)\leq\dots\leq\mathbf{x}_{[n]}(t)$.
(ii) The ordered average opinions among the   clique   $\mathfrak{N}$ are defined as  $\bar{\mathbf{x}}_{[1]}(t),\bar{\mathbf{x}}_{[2]}(t),\dots,\bar{\mathbf{x}}_{[z]}(t)$, where $\bar{\mathbf{x}}_{[k]}(t)$ represents the $k$'th smallest average node states among the ${n}\choose{m} $ cliques in $\mathfrak{N}$.
\end{definition}

The next theorem further shows that the node states may even preserve their orders throughout the entirety of the time horizon  under certain range of initial values. Denote $\Delta_k:=\bar{\mathbf{x}}_{[1]}(0)-\mathbf{x}_{[k]}(0)$.

\begin{theorem}\label{cor-propor-1}
Assume  $m=n$ and  $0\leq\Delta_k\leq\eta$ for some $k\in \mathrm{V}$. Further let the following hold:
$$
\min_{s\in\mathrm{V}/\{k\}}|\mathbf{x}_{[s]}(0)-\bar{\mathbf{x}}_{[1]}(0)|>\eta+\frac{\Delta_k}{n-1}.
$$
Then along the opinion dynamics (\ref{mode}), the order of the  node states $x_i(t)$ is preserved for all  $t=0,1,\dots$ and for all $i\in\mathrm{V}$. In this case, there holds almost surely that $$\lim_{t\to\infty}\mathbf{x}_{[k]}(t)=\lim_{t\to\infty}\bar{\mathbf{x}}_{[1]}(t)=\bar{\mathbf{x}}_{[1]}(0)+\frac{\Delta_k}{n-1}. $$
\end{theorem}

We remark that the condition $m=n$ in Theorems \ref{allselected-converge}, \ref{thm2}, \ref{cor-propor-1}  in social network context implies the clique opinion is from the entire network. In other words, peers are under the pressure of the society's average opinion at any iteration step. We believe similar results would continue to hold for general $m$, evident from Example 1 where $m=4$ and $n=20$, and the proof can be established by extending the same line of analysis for $m=n$. However, a full treatment for that would be much more involved as the proof relies on explicit construction of certain subtle probabilistic events.

\subsection{Opinion Fluctuations}
We introduce the following definition on the fluctuation events.

\begin{definition} The fluctuation event for the opinion dynamics (\ref{mode}) is defined as
$$
\mathsf{E}_{\rm fluctuation}:=\Big\{\exists i\in\mathrm{V} \textrm{ s.t. } \liminf_{t\to\infty}x_{i}(t)<\limsup_{t\to\infty}x_{i}(t)\Big\}.
$$
\end{definition}

We present the following theorem which establishes a condition under which the fluctuation takes places with a strictly positive probability along  (\ref{mode}).
\begin{theorem}\label{Lower-fluctuation}
Let $m\geq 4$ and $n<2m-1$. Suppose $\eta<{1}/{(2m+\frac{2m^{3}}{m-1})}$.  Then along  the opinion dynamics (\ref{mode}) $\mathsf{E}_{\rm fluctuation}$ is a nontrivial event.
\end{theorem}

From the proof of Theorem \ref{Lower-fluctuation}, a lower bound of $\mathbb{P}(\mathsf{E}_{\rm fluctuation})$ may be established explicitly.  It is of interest to further have clear sets of initial values under which $\mathsf{E}_{\rm fluctuation}$ can be proven to happen. To this end, we constructively define the following event of the initial node states:
$$
\begin{array}{rl}
\mathsf{E}_{K}^0:=\Big\{x_{1}(0)=\dots=x_{K-1}(0)=0,
x_{K}(0)\in(\frac{1}{2}-\beta,\frac{1}{2}+\beta),
x_{K+1}(0)=\dots=x_{n}(0)=1\Big\}.
\end{array}
$$
We present the following result.

\begin{theorem} \label{fluctuation-special-initial}
Let ${n+3}/{2}\leq m\leq{2n}/{3}$ and $n-m+2\leq K\leq m-1$. Suppose $\beta<\eta$   and $\eta< {1}/{(6+\frac{4n}{m(n-1)})}$. Then along the opinion dynamics (\ref{mode}) there holds
$$
\mathbb{P}\Big( \mathsf{E}_{\rm fluctuation} \Big| \mathsf{E}_{K}^0 \Big)=1.
$$
\end{theorem}
\subsection{Further Discussions}
The proofs for the  results stated in this section are available in the appendix. Most of the proofs are established by constructive arguments, where we carry out probabilistic analysis on a series of special events for the initial values. Such events are established based on the ordered statistics of the node states and the ordered statistics of the clique  average states. Then the effects of the bounded confidence  in the convergence or fluctuation events are estimated with upper and/or lower bounds, which eventually lead to the presented  results. Compared to classical Deffuant-Weisbuch and Hegselmann-Krause models and their  variations, the clique bounded confidence  opinion dynamics (\ref{mode}) brings in the interplay between the size of the network $n$ and the size of the cliques  $m$. Such an interplay, uncovers the new phenomena that for both consensus and disagreement may happen with nontrivial probabilities, e.g., Theorem 1 and Theorem 2; and  fluctuations also take place with a nontrivial probability which may be determined entirely from the initial states, e.g., Theorem 4 and Theorem 5. To the best of our knowledge, these kinds of results are established for the first time in the literature for bounded confidence models. In the meantime, the coupling between $n$ and $m$ brings fundamental difficulties in the  analysis, which largely limited our study to a few special ranges of the parameters $(n,m,\eta,\delta)$  and node initial states.

\section{Conclusions}\label{sec5}
We have proposed a generalized Deffuant-Weisbuch   model where  the evolution of opinions over a social network  is governed by  bounded confidence in social cliques.  With initial opinions being independently and identically distributed,  at each time step, peers review the average opinions of a randomly selected local clique with a prescribed cardinality. Then nodes update their opinions by averaging their current opinion and the randomly realized clique  average, only when the clique  average is within the bounded confidence intervals. We proved a series of results on the asymptotic behaviors of the social opinions at a system level, focusing on three events: consensus, disagreement and fluctuations. Surprisingly, all three events would happen under certain nontrivial probabilities for given network conditions, in sharp contrast of the universal clustering behavior in bounded confidence social network models. Future works include extending the results to general network structures, and validations of the established opinion formations with real-world social network data.

\section*{Appendix}

\section*{A. Proof of Theorem 1}
Before proceeding to giving explicit proofs of Theorem 1, some fundamental preliminaries on the initial node states are presented below.

Denote the network node states $\mathbf{X}(t)=(x_{1}(t),x_{2}(t),\ldots,x_{n}(t))^\top$ for all $t\geq 0$. With    $\bar{\mathbf{x}}_{[k]}(t)$ defined as the $k$'th smallest average node states among the ${n}\choose{m}$ cliques  in $\mathfrak{N}$, we introduce three disjoint sets for the initial state $\mathbf{X}(0)$ as below.

\begin{itemize}
\item[(i)]$\mathcal{I}_{1}=\bigcup_{k=1}^{n-1}\mathcal{A}_{k}$ with
\[\begin{array}{l}
\mathcal{A}_{k}:=\Big\{\mathbf{X}(0):  |\bar{\mathbf{x}}_{[1]}(0)-x_{i}(0)|\leq\eta, i=1,2,\dots,k; \\    \qquad\qquad\qquad \min\limits_{j\geq k+1}|x_{j}(0)-\bar{\mathbf{x}}_{[1]}(0)|\geq\eta+\frac{1}{n-k}\sum\limits_{i=1}^{k}|\bar{\mathbf{x}}_{[1]}(0)-x_{i}(0)| \Big\}
\end{array}\]
\item[(ii)] $\mathcal{I}_{2}=\bigcup_{k=1}^{n-2}\bigcup_{l=1}^{n-k}\mathcal{B}_{kl}$ with
\[\begin{array}{ll}
\mathcal{B}_{kl}:=\Big\{ &\mathbf{X}(0):  |\bar{\mathbf{x}}_{[1]}(0)-x_{i}(0)|\leq\eta, i=1,2,\dots,k;\\
&|x_{j}(0)-\bar{\mathbf{x}}_{[1]}(0)|\in(\eta,\eta+\frac{1}{n-k}\sum\limits_{s=1}^{k}|\bar{\mathbf{x}}_{[1]}(0)-x_{s}(0)|), j=k+1,k+2,\dots,k+l;\\
&\min\limits_{j\geq k+l+1}\{|x_{j}(0)-\bar{\mathbf{x}}_{[1]}(0)|\}\geq \eta +\frac{1}{n-k}\sum\limits_{s=1}^{k}|\bar{\mathbf{x}}_{[1]}(0)-x_{s}(0)| \Big\}
\end{array}\]
\item[(iii)]$\mathcal{I}_{3}=\mathcal{C}_1 \bigcup \mathcal{C}_2$ with
    \[\begin{array}{l}
    \mathcal{C}_1 := \Big\{\mathbf{X}(0):\ \min_{i=1,2,\dots,n}|x_{i}(0)-\bar{\mathbf{x}}_{[1]}(0)|>\eta \Big\}\,\\
    \mathcal{C}_2 := \Big\{\mathbf{X}(0):\ \max_{i=1,2,\dots,n}|x_{i}(0)-\bar{\mathbf{x}}_{[1]}(0)|\leq\eta\Big\}\,.
    \end{array}\]
\end{itemize}
It is worth noting that each $\mathcal{I}_{i}$, $i=1,2,3$ is defined as a union of several disjoint sets, and given any initial network state,  there is a unique ${i}\in\{1,2,3\}$ such that the corresponding initial network state $\mathbf{X}(0)\in\mathcal{I}_{i}$ by appropriately ordering the network nodes.

In view of the above analysis,  the proof reduces to show that all node states converge to a finite limit under $\mathbf{X}(0)\in\mathcal{I}_{i}$, for each $i=1,2,3$.  Note that, when $m=n$, the updates of the $x_i(t)$ for all $i\in \mathrm{V}$  becomes deterministic once the initial values are randomly assigned, and $\bar{\mathbf{x}}_{[k]}(t)=\sum_{i=1}^{n}x_i(t)/n$ for all $k\in\mathrm{V}$ and $t\geq 0$.

\subsection*{A.1 Proof for $\mathbf{X}(0)\in\mathcal{I}_{1}$}

Note that  $\mathcal{I}_{1}$ is a union of disjoint subsets $\mathcal{A}_{k}$, $k=1,2,\dots,n-1$. Thus, in this subsection we approach the proof by studying the convergence of the update rule (\ref{mode}) under $\mathbf{X}(0)\in\mathcal{A}_{k}$  for each $k=1,2,\dots,n-1$.

First of all, we consider $\mathbf{X}(0)\in\mathcal{A}_{1}$, and proceed to show that the node state $x_1(t)$ converges a.s. and all other node states $x_i(t)$, $i=2,\ldots,n$ remain as their initial values for all $t\geq0$. Without loss of generality, we assume that $0<\bar{\mathbf{x}}_{[1]}(0)-x_{1}(0)<\eta$.

When $t=1$, it immediately follows from (\ref{mode}) that $x_{i}(1)=x_{i}(0)$ for all $i\neq 1$, and
\begin{align}\label{e:convergence1}
\left\{
\begin{array}{ll}
x_{1}(1)=x_{1}(0)+\delta(\bar{\mathbf{x}}_{[1]}(0)-x_{1}(0))>x_{1}(0)\\
\bar{\mathbf{x}}_{[1]}(1)=\frac{1}{n}\sum\limits_{j=1}^{n}x_{j}(1)=\bar{\mathbf{x}}_{[1]}(0)+\frac{\delta}{n}(\bar{\mathbf{x}}_{[1]}(0)-x_{1}(0))\,.
\end{array}
\right.
\end{align}
Besides, we can also establish
\begin{align}
&\min_{j\neq 1}|x_{j}(1)-\bar{\mathbf{x}}_{[1]}(1)| \nonumber\\
&\stackrel{a)}{\geq}\eta+\frac{1}{n-1}(\bar{\mathbf{x}}_{[1]}(0)-x_{1}(0))-\frac{\delta}{n}(\bar{\mathbf{x}}_{[1]}(0)-x_{1}(0)) \nonumber\\
&\stackrel{b)}{>}\eta \label{e:convergence2}
\end{align}
where $a)$ is obtained by using (\ref{e:convergence1}) and $b)$ is obtained by using $\frac{1}{n-1}>\frac{\delta}{n}$. Besides, with (\ref{e:convergence1}), it follows that
\[\begin{array}{rcl}
|\bar{\mathbf{x}}_{[1]}(1)-x_{1}(1)|
&=&(1-\delta+\frac{\delta}{n})|\bar{\mathbf{x}}_{[1]}(0)-x_{1}(0)|\\
&\leq& (1-\delta+\frac{\delta}{n})\eta\\
&<&\eta.
\end{array}\]

When $t=2$, similarly we can obtain from (\ref{mode}) that $x_{i}(2)=x_{i}(1)$ for all $i\neq 1$, and
\[\left\{
\begin{array}{l}
x_{1}(2)=x_{1}(1)+\delta(1-\delta+\frac{\delta}{n})(\bar{\mathbf{x}}_{[1]}(0)-x_{1}(0))>x_{1}(1)\\
\bar{\mathbf{x}}_{[1]}(2)=\bar{\mathbf{x}}_{[1]}(0) +\frac{\delta}{n}(2-\delta+\frac{\delta}{n})
(\bar{\mathbf{x}}_{[1]}(0)-x_{1}(0)).
\end{array}
\right.\]
Besides, we can also establish
\begin{align*}
&\min_{j\neq 1}|x_{j}(2)-\bar{\mathbf{x}}_{[1]}(2)|=\min_{j\neq 1}|x_{j}(0)-\bar{\mathbf{x}}_{[1]}(2)|\\
&=\min_{j\neq 1}\left|x_{j}(0)-\bar{\mathbf{x}}_{[1]}(0)-\frac{\delta}{n}\big(2-\delta+\frac{\delta}{n}\big)
(\bar{\mathbf{x}}_{[1]}(0)-x_{1}(0))\right|\\
&{\geq}\min_{j\neq1}|x_{j}(0)-\bar{\mathbf{x}}_{[1]}(0)|-\frac{\delta}{n}\big(2-\delta+\frac{\delta}{n}\big)|\bar{\mathbf{x}}_{[1]}(0)-x_{1}(0)|\\
&\stackrel{a)}{\geq}\eta+\left(\frac{1}{n-1}-\frac{\delta}{n}(2-\delta+\frac{\delta}{n})\right)|\bar{\mathbf{x}}_{[1]}(0)-x_{1}(0)|>\eta
\end{align*}
and
\begin{align*}
|\bar{\mathbf{x}}_{[1]}(2)-x_{1}(2)|
&=(1-\delta+\frac{\delta}{n})^2|\bar{\mathbf{x}}_{[1]}(0)-x_{1}(0)|\\
&\leq (1-\delta+\frac{\delta}{n})^2\eta\\
&<\eta\,,
\end{align*}
where $a)$  comes from (\ref{e:convergence1}) and (\ref{e:convergence2}).

Along this way, we can recursively apply the previous arguments to the case when $t=3,4,\ldots$. For any $t\geq 3$, suppose $x_{i}(t)=x_{i}(0)$ for all $i\neq 1$ and
\begin{align}\label{E:lemB}
\left\{
\begin{array}{l}
x_{1}(t)=x_{1}(t-1)+\delta(1-\delta+\frac{\delta}{n})^{t-1}(\bar{\mathbf{x}}_{[1]}(0)-x_{1}(0))\, >x_{1}(t-1)\\
\bar{\mathbf{x}}_{[1]}(t)=\bar{\mathbf{x}}_{[1]}(0)+\frac{\delta}{n}\sum\limits_{j=0}^{t-1}\left(1-\delta+\frac{\delta}{n}\right)^{j}
(\bar{\mathbf{x}}_{[1]}(0)-x_{1}(0))
\end{array}
\right.
\end{align}
where
$|\bar{\mathbf{x}}_{[1]}(t)-x_{1}(t)|\leq\eta$ and $\min_{j\neq 1}|x_{j}(t)-\bar{\mathbf{x}}_{[1]}(t)|>\eta$. This, together with (\ref{mode}), leads to
$$
x_{j}(t+1)=x_{j}(t)=x_{j}(0), \ \forall j\neq 1\,
$$
and
\begin{align}\label{E:lemB}
\begin{array}{ll}
x_{1}(t+1)=x_{1}(t)+\delta(1-\delta+\frac{\delta}{n})^{t}(\bar{\mathbf{x}}_{[1]}(0)-x_{1}(0))>x_{1}(t)\\
\bar{\mathbf{x}}_{[1]}(t+1)=\bar{\mathbf{x}}_{[1]}(0)+\frac{\delta}{n}\sum\limits_{j=0}^{t}\left(1-\delta+\frac{\delta}{n}\right)^{j}
(\bar{\mathbf{x}}_{[1]}(0)-x_{1}(0)).
\end{array}
\end{align}
Furthermore, by some simple calculations, it can be concluded that $|\bar{\mathbf{x}}_{[1]}(t+1)-x_{1}(t+1)|\leq\eta$ and $\min_{j\neq 1}|x_{j}(t+1)-\bar{\mathbf{x}}_{[1]}(t+1)|>\eta$.

\medskip
In summary of the previous analysis, we can conclude that $x_{j}(t)=x_{j}(0)$ for all $j\neq 1$ and $t\geq 0$,  and
\[
x_1(t) = x_1(0) + \sum_{k=0}^{t-1}\delta(1-\delta+\frac{\delta}{n})^{k}(\bar{\mathbf{x}}_{[1]}(0)-x_{1}(0))\,,
\]
yielding
\[\begin{array}{rcl}
\lim_{t\rightarrow\infty}
x_{1}(t)&=&x_1(0) + \frac{n}{n-1}(\bar{\mathbf{x}}_{[1]}(0)-x_{1}(0))\\
&=&\frac{1}{n-1}\sum_{j\neq 1}x_j(0)\,.
\end{array}\]

In the previous analysis, we have shown that all node states asymptotically converges to some finite values under the initial conditions $\mathbf{X}(0)\in\mathcal{A}_{1}$.
Similarly, we can apply the previous arguments to other initial conditions $\mathbf{X}(0)\in\mathcal{A}_{k}$, $k=2,\ldots,n-1$, leading to an asymptotic convergence of all node states. For the sake of simplicity, the corresponding details is omitted.
Therefore, the statements in Theorem 1 can be concluded for the initial states $\mathbf{X}(0)\in\mathcal{I}_{1}$.

\subsection*{A.2 Proof for  $\mathbf{X}(0)\in\mathcal{I}_{2}$}

Note that the set $\mathcal{I}_{2}$ is comprised of a number of disjoint subsets $\mathcal{B}_{kl}$.  We now proceed to prove the theorem for each subset $\mathcal{B}_{kl}$.

We first consider the pair $(k,l)=(1,1)$, i.e.,
\begin{align}\label{e:lemC}
\left\{
\begin{array}{lll}
0<\bar{\mathbf{x}}_{[1]}(0)-x_{1}(0)\leq\eta\\
\eta<x_{2}(0)-\bar{\mathbf{x}}_{[1]}(0)<\eta+\frac{1}{n-1}(\bar{\mathbf{x}}_{[1]}(0)-x_{1}(0))\\
\min_{j\geq 3}\{|x_{j}(0)-\bar{\mathbf{x}}_{[1]}(0)|\}\geq\eta+\frac{1}{n-1}(\bar{\mathbf{x}}_{[1]}(0)-x_{1}(0)).
\end{array}
\right.
\end{align}
Similar to the previous analysis for the  case $\mathbf{X}(0)\in\mathcal{I}_{1}$, we can establish that $x_{1}(t)$ and $\bar{\mathbf{x}}_{[1]}(t)$ increase until $x_{2}(0)-\bar{\mathbf{x}}_{[1]}(t^{*})\leq\eta$ for certain threshold $t^{*}>0$. Particularly, we have
\begin{align*}
x_{1}(t)=x_{1}(t-1)+\delta(\bar{\mathbf{x}}_{[1]}(t-1)-x_{1}(t-1))>x_{1}(t-1)
\end{align*}
where
\begin{align}\label{e:lemD}
\bar{\mathbf{x}}_{[1]}(t)-x_{1}(t)=(1-\delta+\frac{\delta}{n})^{t}(\bar{\mathbf{x}}_{[1]}(0)-x_{1}(0))>0
\end{align}
for $t=1,2,\ldots,t^{*}$, and $x_{j}(t)=x_j(0)$ for all $j\neq 1$ and $t=1,2,\ldots,t^*-1$, with $t^*$ being such that
\[\begin{array}{ll}
&x_{2}(t^*)-\bar{\mathbf{x}}_{[1]}(t^*)\\
\stackrel{a)}{=}&(x_{2}(0)-\bar{\mathbf{x}}_{[1]}(0))-\frac{1-(1-\delta+\frac{\delta}{n})^{t^*}}{n-1}(\bar{\mathbf{x}}_{[1]}(0)-x_{1}(0))\\
\stackrel{b)}{\leq}&\eta
\end{array}\]
where $a)$ is based on (\ref{E:lemB}) and $b)$ comes from (\ref{e:lemC}). Note that
\[\begin{array}{ll}
&x_{2}(t^*-1)-\bar{\mathbf{x}}_{[1]}(t^*-1)\\
=&(x_{2}(0)-\bar{\mathbf{x}}_{[1]}(0))-\frac{1-(1-\delta+\frac{\delta}{n})^{t^{*}-1}}{n-1}(\bar{\mathbf{x}}_{[1]}(0)-x_{1}(0))\\
>&\eta
\end{array}\]
and
\[\begin{array}{ll}
&x_{2}(t^*)-\bar{\mathbf{x}}_{[1]}(t^*)\\
=&(x_{2}(0)-\bar{\mathbf{x}}_{[1]}(0))-\frac{1-(1-\delta+\frac{\delta}{n})^{t^*}}{n-1}(\bar{\mathbf{x}}_{[1]}(0)-x_{1}(0))\\
\leq&\eta.
\end{array}\]
Thus, we can calculate the threshold
$$t^{*}=\bigg\lceil\frac{\ln{(1-(n-1)\frac{x_{2}(0)-\bar{\mathbf{x}}_{[1]}(0)-\eta}{\bar{\mathbf{x}}_{[1]}(0)-x_{1}(0)})}}{\ln{(1-\delta+\frac{\delta}{n})}}\bigg\rceil.$$

At time $t^{*}+1$, we have
\begin{align*}
\left\{
\begin{array}{ll}
x_{i}(t^{*}+1)=x_{i}(t^{*})+\delta(\bar{\mathbf{x}}_{[1]}(t^{*})-x_{i}(t^{*})),i=1,2\\
\bar{\mathbf{x}}_{[1]}(t^{*}+1)=\bar{\mathbf{x}}_{[1]}(t^{*})+\frac{\delta}{n}\sum_{j=1}^{2}(\bar{\mathbf{x}}_{[1]}(t^{*})-x_{j}(t^{*})),
\end{array}
\right.
\end{align*}
which implies that for $t=t^*$ there holds
\begin{align*}
\sum_{j=1}^{2}(\bar{\mathbf{x}}_{[1]}(t+1)-x_{j}(t+1))
=(1-\delta+\frac{2\delta}{n})\sum_{j=1}^{2}(\bar{\mathbf{x}}_{[1]}(t)-x_{j}(t))\,.
\end{align*}
Moreover, when $t>t^{*}$, it can be verified that $x_{i}(t)=x_i(0)$ for $i=3,4,\ldots,n$, and for $i=1,2$
\begin{align*}
x_{i}(t)=(1-\delta)^{t-t^{*}}x_{i}(t^{*})+\sum_{k=0}^{t-t^{*}-1}\delta(1-\delta)^{k}\bar{\mathbf{x}}_{[1]}(t-1-k),
\end{align*}
which yields
\begin{align*}
&x_{1}(t)-x_{1}(t-1)=\delta(1-\delta)^{t-t^{*}-1}(\bar{\mathbf{x}}_{[1]}(t^{*})-x_{1}(t^{*}))\\
&x_{2}(t)-x_{2}(t-1)=\delta(1-\delta)^{t-t^{*}-1}(\bar{\mathbf{x}}_{[1]}(t^{*})-x_{2}(t^{*}))\,.
\end{align*}
Therefore, all node states $x_{j}(t)$, $j\in\mathrm{V}$ are convergent a.s..

\medskip
In the previous analysis, we have shown that all node states converge to some finite values almost surely under the initial conditions $\mathbf{X}(0)\in\mathcal{B}_{kl}$ with $(k,l)=(1,1)$. Similarly,  we can establish the same conclusions under other initial conditions $\mathbf{X}(0)\in\mathcal{B}_{kl}$ for other pairs of $(k,l)$, which is omitted herein for the sake of simplicity.  Therefore, the statements in Theorem 1 are shown to be true for all initial states $\mathbf{X}(0)\in\mathcal{I}_2$.

\subsection*{A.3 Proof for  $\mathbf{X}(0)\in\mathcal{I}_{3}$}

For the initial value set $\mathcal{I}_{3}:=\mathcal{C}_1 \bigcup \mathcal{C}_2$, we can draw the following immediate conclusions.
\begin{itemize}
\item[(i).] When $\mathbf{X}(0)\in\mathcal{C}_1$, there  holds $x_i(t)=x_i(0)$ for all $i\in\mathrm{V}$ and $t\geq0$.
\item[(ii).] When $\mathbf{X}(0)\in\mathcal{C}_2$, the node state updating rule (\ref{mode}) becomes a standard DeGroot model over a complete interaction graph. Thus, all node states $x_i(t)$ converge to the average of the initial states.
\end{itemize}
The desired theorem holds.

\section*{B Proof of Theorem 2}

Denote the state range at time $t$ as $D_{[i,j]}(t)=x_{[j]}(t)-x_{[i]}(t)$ for all $t\geq 0$ and $i,j\in\mathrm{V}$. We then introduce three disjoint sets for the initial state $\mathbf{X}(0)$ as below.
\begin{itemize}
\item[(i)]$\mathcal{I}^\ast_{1}=\Big\{\mathbf{X}(0):D_{[1,n]}(0)<\eta \Big\}$;
\item[(ii)] $\mathcal{I}^\ast_{2}=\Big\{ \mathbf{X}(0): x_{1}(0)=0,x_{i}(0)\in(1-\eta^n,1],i=2,3,\ldots,n \Big\}$;
\item[(iii)]$\mathcal{I}^\ast_{3}=\Big\{\mathbf{X}(0):x_{k}(0) \textrm{ are pairwise different}, k\in\mathrm{V}, x_{i}(0)\in[0,\eta),i=1,2,\ldots,\lfloor\frac{n}{2}\rfloor,x_{j}(0)\in(1-\eta,1],j=\lfloor\frac{n}{2}\rfloor+1,\ldots,n \Big\}$.
\end{itemize}

With these sets being the case, the proof can be divided into  two steps, corresponding to the subsequent two subsections, respectively.

\subsection*{B.1 Non-triviality of $\mathsf{E}_{\rm consensus}$}

First of all, we note that all opinions reach consensus if $D_{[1,n]}(0)<\eta$. Thus, the limit state set for $\mathbf{X}(0)\in\mathcal{I}^\ast_{1}$ is a subset of $\mathsf{E}_{\rm consensus}$. Besides, $\mathbb{P}\{\mathcal{I}^\ast_{1}\}>0$, yielding that $\mathbb{P}\{\mathsf{E}_{\rm consensus}\}>0$.

Next we proceed to prove that $\mathbb{P}\{\mathsf{E}_{\rm consensus}\}<1$ by showing that the node states under $\mathbf{X}(0)\in\mathcal{I}^\ast_{2}$ do not reach consensus.
Given any $\mathbf{X}(0)\in\mathcal{I}^\ast_{2}$, it can be found that
\[
\frac{(n-1)(1-\eta^n)}{n}<\bar{\mathbf{x}}_{[1]}(0)\leq 1-\frac{1}{n}<x_{i}(0), \quad \forall i=2,3,\ldots,n.
\]
This then implies
\begin{align*}
\left\{
\begin{array}{ll}
\bar{\mathbf{x}}_{[1]}(0)-x_{1}(0)>\frac{(n-1)(1-\eta^n)}{n}>\eta\\
x_{i}(0)-\bar{\mathbf{x}}_{[1]}(0)>1-\eta^{n}-(1-\frac{1}{n})\stackrel{a)}{>}\eta, \quad i=2,3,\ldots,n
\end{array}
\right.
\end{align*}
where we have used $\eta^{n-1}<(\frac{1}{n+1})^{n-1}<\frac{1}{n}$ to obtain $a)$. Thus, all node states $x_i(t)$ remain unchanged for all $t$. Therefore, we have $\mathbb{P}\{\mathcal{I}^\ast_{2}\}>0$ and  $\mathbb{P}\{\mathsf{E}_{\rm consensus}\}<1$.

\subsection*{B.2 Non-triviality of $\mathsf{E}_{\rm disagreement}$}

It is observed that $\mathsf{E}_{\rm consensus}\cap\mathsf{E}_{\rm disagreement}=\emptyset$, which indicates $\mathbb{P}\{\mathsf{E}_{\rm disagreement}\}\leq 1-\mathbb{P}\{\mathsf{E}_{\rm consensus}\}<1$ by $0<\mathbb{P}\{\mathsf{E}_{\rm consensus}\}<1$.
With this in mind, we now proceed to  prove that the limit set of $\mathcal{I}^\ast_{3}$ is a subset of $\mathsf{E}_{\rm disagreement}$.

We note that
\begin{align*}
\begin{array}{rcl}
&&(\bar{\mathbf{x}}_{[1]}(0)-\eta,\bar{\mathbf{x}}_{[1]}(0)+\eta)\\
&\subset&\large[(1-\eta)(1-\lfloor\frac{n}{2}\rfloor\frac{1}{n})-\eta,\,1-\frac{1-\eta}{n}\lfloor\frac{n}{2}\rfloor+\eta\large]\\
&\stackrel{a)}{\subset}&(\frac{1-3\eta}{2},\frac{1+3\eta}{2})\\
&\stackrel{b)}{\subset}&(\eta,1-\eta)
\end{array}\end{align*}
where $a)$ is obtained by using  $\lfloor\frac{n}{2}\rfloor\frac{1}{n}\leq\frac{1}{2}$ and $b)$ is obtained by using $\eta\leq\frac{1}{5}\leq\frac{1}{n+1}$ and $\min_{i\in\mathrm{V}}|x_{i}(0)-\bar{\mathbf{x}}_{[1]}(0)|>\eta$. It then immediately follows that  $\textbf{X}(t)=\textbf{X}(0)$ for $\textbf{X}(0)\in\mathcal{I}^\ast_{3}$. Thus, the limit set of $\mathcal{I}^\ast_{3}$ is a subset of $\mathsf{E}_{\rm disagreement}$, i.e., $\mathbb{P}\{\mathcal{I}^\ast_{3}\}>0$ and $\mathbb{P}\{\mathsf{E}_{\rm disagreement}\}>0$. This proves the non-triviality of $\mathsf{E}_{\rm disagreement}$.

The proof is thus completed.

\section*{C Proof of Theorem 3}

\subsection*{C.1 Order Preservation}

In this subsection, we aim to prove that along the opinion dynamics (\ref{mode}), the order of the node states $x_i(t)$ is preserved for all  $t=0,1,\dots$ and for all $i\in\mathrm{V}$. {We fix any $k$ and denote $x_{i}(0)\triangleq \mathbf{x}_{[i]}(0)$ for $i\in\mathrm{V}$.
Then, with $|x_{k}(0)-\bar{\mathbf{x}}_{[1]}(0)|\leq\eta$  and  $|x_{i}(0)-\bar{\mathbf{x}}_{[1]}(0)|>\eta$ for $i\neq k$}, we have
$$\begin{array}{rcl}
x_{k}(1)-x_{i}(1)&=&(1-\delta)(x_{k}(0)+\delta\bar{\mathbf{x}}_{[1]}(0)-x_{i}(0))\\
&\geq& x_{k}(0)-x_{i}(0)\\
&>& 0
\end{array}$$
for $1\leq i< k$ and
$$\begin{array}{rcl}
x_{j}(1)-x_{k}(1)&=&x_{j}(0)-(1-\delta)(x_{k}(0)-\delta\bar{\mathbf{x}}_{[1]}(0)\\
&\geq& x_{j}(0)-\bar{\mathbf{x}}_{[1]}(0)\\
&>& 0
\end{array}$$
for $k<j\leq n$. Thus, the order of the node states $x_i(t)$ is preserved at $t=1$.

For any $t\geq1$, we suppose that $x_{i}(t)\leq x_{k}(t)\leq x_{j}(t)$ for $i<k<j$ and the order is preserved. It can be seen from the proof of Theorem 1 that $|x_{k}(t)-\bar{\mathbf{x}}_{[1]}(t)|\leq\eta$. Then, according to (\ref{E:lemB}), node states $x_{k}(t)$ and the average $\bar{\mathbf{x}}_{[1]}(t)$ strictly increase, which yields
$$\begin{array}{rcl}
x_{k}(t+1)-x_{i}(t+1)&=&(1-\delta)(x_{k}(t)+\delta\bar{\mathbf{x}}_{[1]}(t)-x_{i}(0))\\
&\geq& x_{k}(0)-x_{i}(0)\\
&>& 0
\end{array}$$
for $1\leq i< k$ and
$$\begin{array}{rcl}
x_{j}(t+1)-x_{k}(t+1)&=&x_{j}(0)-(1-\delta)(x_{k}(t)-\delta\bar{\mathbf{x}}_{[1]}(t)\\
&\geq& x_{j}(0)-\bar{\mathbf{x}}_{[1]}(t)\\
&>& 0
\end{array}$$
for $k<j\leq n$. Thus, the opinion order is preserved at time $t+1$.

\subsection*{C.2 Convergence Limits}

Recalling the analysis in Appendix A.1 for $\mathbf{X}(0)\in\mathcal{I}_{1}$, we can obtain
\begin{align*}
\begin{array}{rcl}
x_{k}(1)&=&x_{k}(0)+\delta\Delta_k\\
x_{j}(1)&=&x_{j}(0), \qquad j\neq k\\
\bar{\mathbf{x}}_{[1]}(1)&=&\bar{\mathbf{x}}_{[1]}(0)+\frac{\delta}{n}\Delta_k.
\end{array}\end{align*}
This, together with (\ref{mode}), implies
\begin{align*}
&x_{k}(2)=x_{k}(1)+\delta(\bar{\mathbf{x}}_{[1]}-x_{k}(1))=x_{k}(1)+\delta(1-\delta+\frac{\delta}{n})\Delta_k \\ &x_{j}(2)=x_{j}(0), \qquad\qquad j\neq k\\ &\bar{\mathbf{x}}_{[1]}(2)=\bar{\mathbf{x}}_{[1]}(0)+\frac{\delta}{n}\left(1+(1-\delta+\frac{\delta}{n})\right)\Delta_k.
\end{align*}
Furthermore, by induction we can obtain
\begin{align*}
&x_{k}(t)=x_{k}(0)+\delta\sum_{k=0}^{t-1}\left(1-\delta+\frac{\delta}{n}\right)^{k}\Delta_k, \\ &x_{j}(t)=x_{j}(0), \qquad j\neq k;\\
&\bar{\mathbf{x}}_{[1]}(t)=\bar{\mathbf{x}}_{[1]}(0)+\frac{\delta}{n}\sum_{k=0}^{t-1}\left(1-\delta+\frac{\delta}{n}\right)^{k}\Delta_k.
\end{align*}
The first and third of the above equations immediately render that
\begin{align*}
&\lim_{t\to\infty}x_{k}(t)=x_{k}(0)+\frac{n}{n-1}\Delta_k=\bar{\mathbf{x}}_{[1]}(0)+\frac{\Delta_k}{n-1}.\\
&\lim_{t\to\infty}\bar{\mathbf{x}}_{[1]}(t)=\bar{\mathbf{x}}_{[1]}(0)+\frac{\Delta_k}{n-1}
\end{align*}

The proof is thus completed.

\section*{D. Proof of Theorem 4}

Before proceeding to the explicit proofs, we first introduce some instrumental terminologies to facilitate the subsequent analysis.
We fix the agent indexes by letting $x_{i}(0)\triangleq \mathbf{x}_{[i]}(0)$ for $i\in\mathrm{V}$ and let $s\in\{n-m+1,\dots,m-1\}$.
Denote $\mathcal{G}_{1}=\{1,2,\dots,s-1\}$, $\mathcal{G}_{2}=\{s\}$ and $\mathcal{G}_{3}=\{s+1,\dots,n\}$, and by $\mathcal{S}_{i}(0)=(k_{1}^{(i)}(0),k_{2}^{(i)}(0),k_{3}^{(i)}(0))$ the selection tube of the average opinion $\bar{\mathbf{x}}_{i}(0)$, $i\in\bar{\mathrm{V}}=\{1,2,\dots,C_{n}^{m}\}$ where there are $k_{k}^{(i)}(0)$ opinions selected from $\mathcal{G}_{k}$, $k=1,2,3$ for the average opinion $\bar{\mathbf{x}}_{i}(0)$ at time $0$. For simplicity, we denote $\bar{\mathbf{x}}_{i}(0)\in\mathcal{S}_{i}(0)=(k_{1}^{(i)}(0),k_{2}^{(i)}(0),k_{3}^{(i)}(0))$\footnote{Particularly, for the selection tube, $\sum_{s=1}^{3}k_{s}^{(i)}(0)=m,\textrm{ }k_{2}^{(i)}(0)=0 \textrm{ or }1$ for any $i\in\bar{\mathrm{V}}$.}. Define an equivalent relation $\tilde{\mathcal{S}}$ for the average opinion $\{\bar{\mathbf{x}}_{i}(0)\}$, that is, if $\mathcal{S}_{i}(0)=\mathcal{S}_{j}(0), i\in\bar{\mathrm{V}}$, then $\bar{\mathbf{x}}_{i}(0)\tilde{\mathcal{S}}\bar{\mathbf{x}}_{j}(0)$. Then the quotient set of the average agent index set $\bar{\mathrm{V}}$ on the equivalent relation $\tilde{\mathcal{S}}$ is defined  as $$\mathcal{H}=\bar{\mathrm{V}}/\tilde{\mathcal{S}}=\{\mathcal{H}_{l}, l=1,2,\dots,K_{s}\},$$ where $K_{s}=\min\{s,n-s+1,m+1\}+\min\{s,n-s+1,m\}$\footnote{Note that there are at most $s-1$ agents in $\mathcal{G}_{1}$ and $n-s$ agents in $\mathcal{G}_{3}$. Thus, there are $\min\{s,n-s+1,m+1\}$ selection methods for $k_{2}^{(i)}(0)=0$. Similarly, there are at most $\min\{s,n-s+1,m\}$ selection methods for $k_{2}^{(i)}(0)=1$. Hence,  there are $K_{s}$ methods in total.}, and
\begin{align*}
\mathcal{H}_{l}=\left\{
\begin{array}{ll}
\{i:\bar{\mathbf{x}}_{i}(0)\in(s-\lceil\frac{l}{2}\rceil,1,m-s+\lceil\frac{l}{2}\rceil-1)\},
\qquad \qquad \,\textrm{ if } l \textrm{ is an even number}\\
\{i:\bar{\mathbf{x}}_{i}(0)\in(s-\lceil\frac{l}{2}\rceil,0,m-s+\lceil\frac{l}{2}\rceil)\},
\qquad \qquad \qquad \textrm{ if } l \textrm{ is an odd number}\,.
\end{array}
\right.
\end{align*}
It can be easily verified that  $\mathcal{S}_{1}(0)=(s-1,1,m-s)$ and  $\mathcal{S}_{C_{n}^{m}}(0)=(m-n+s-1,1,n-s)$.

We denote by $\alpha_{s}=\max\{D_{[1,s-1]}(0),D_{[s+1,n]}(0)\}$  the lower opinion range of agent set $\mathcal{G}_{1}$ and $\beta_{s}=\min\{D_{[s-1,s]}(0),D_{[s,s+1]}(0)\}$ the upper opinion range $\mathcal{G}_{3}$ at $t=0$. Furthermore,  denote
\begin{align*}
&\alpha_{s}^{t}=\max\{D_{[1,s-1]}(t),D_{[s+1,n]}(t)\}\\
&\beta_{s}^{t}=\min\{D_{[s-1,s]}(t),D_{[s,s+1]}(t)\},\quad \\
&\mathcal{R}_{l}=\max_{i,j\in\mathcal{H}_{l}}|\bar{\mathbf{x}}_{[i]}(0)-\bar{\mathbf{x}}_{[j]}(0)|\\
&\mathcal{C}_{l}=\min_{i\notin\mathcal{H}_{l},j\in\mathcal{H}_{l}}|\bar{\mathbf{x}}_{[i]}(0)-\bar{\mathbf{x}}_{[j]}(0)|\\
&\underline{\alpha}_{l}(0)=\min_{i\in\mathcal{H}_{l}}\bar{\mathbf{x}}_{[i]}(0)\\
&\overline{\alpha}_{l}(0)=\max_{i\in\mathcal{H}_{l}}\bar{\mathbf{x}}_{[i]}(0)\,,\quad l\in\{1,\dots,K_{s}\}.
\end{align*}

Now we are ready to present the proof of Theorem 4, consisting of three steps. We first give the parameter condition of opinion fluctuation and the bounds of the parameter ranges in Subsection D.1. Then in Subsection D.2 we show that the opinion fluctuation happens under certain parameter conditions. Finally, we prove that opinion fluctuation happens with a positive probability in Subsection D.3.

\subsection*{D.1 Measures of Quotient Sets}

The following lemma is given to measure the quotient sets of average values $\{\bar{\mathbf{x}}_{[i]}(0),i\in\bar{\mathrm{V}}\}$.

\begin{lemma}\label{lemma-clustering-1}
Given any initial values and $\alpha_{m}>0$,  all average opinion values $\{\bar{\mathbf{x}}_{[i]}(0),i\in\bar{\mathrm{V}}\}$ can be separated into $K_{s}$ cliques, and for any $l\in\{1,2,\dots,K_{s}\}$, there hold
\begin{align}
&\frac{1}{m}\beta_{s}\leq\mathcal{C}_{l}\leq\frac{1}{m}\max\{D_{[s-1,s]}(0),D_{[s,s+1]}\},\label{eq:C_l}\\
&\frac{1}{m}\min\{D_{[1,s-1]}(0),D_{[s+1,n]}(0)\}\leq\mathcal{R}_{l}\leq\alpha_{s},\label{eq:R_l}\\
&
\begin{array}{l}
\frac{1}{m}\min\limits_{i\in\{1,\dots,n-1\}}D_{[i,i+1]}(0)\leq  \max\limits_{i,i+1\in\mathcal{H}_{l}}|\bar{\mathbf{x}}_{[i+1]}(0)-\bar{\mathbf{x}}_{[i]}(0)| \leq\frac{\alpha_{s}}{m}.
\end{array}\label{eq:alpha_l}
\end{align}

\end{lemma}
{\it Proof.}
For any $\{\bar{\mathbf{x}}_{[i]}(0)\}$, $i\in\bar{\mathrm{V}}$, we denote
$$ \bar{\mathbf{x}}_{[i]}(0)=\frac{1}{m}\big(\sum_{l=1,i_{l}\in\mathcal{G}_{1}}^{k_{1}^{(i)}(0)}x_{i_{l}}(0)+k_{2}^{(i)}(0)
x_{s}(0)+\sum_{l=1,i_{l}\in\mathcal{G}_{3}}^{k_{3}^{(i)}(0)}x_{i_{l}}(0)\big).
$$
Then we can obtain that
\begin{equation}\label{lower-fluctuate-1}\begin{array}{l}
\frac{\left(k_{1}^{(i)}(0)x_{1}(0)+k_{2}^{(i)}(0)x_{s}(0)+k_{3}^{(i)}(0)
x_{s+1}(0)\right)}{m}{\leq}\bar{\mathbf{x}}_{[i]}(0)  {\leq}\frac{\left(k_{1}^{(i)}(0)x_{s-1}(0)+k_{2}^{(i)}(0)x_{s}(0)+
k_{3}^{(i)}(0)x_{n}(0)\right)}{m}
\end{array}\end{equation}
where the lower bound is obtained by using $x_{1}(0)\leq x_{i}(0)$ for $i\in\mathcal{G}_{1}$ and $x_{s+1}(0)\leq x_{j}(0)$ for $j\in\mathcal{G}_{3}$, and the upper bound is obtained by using $x_{i}(0)\leq x_{s-1}(0)$ for $i\in\mathcal{G}_{1}$ and $x_{j}(0)\leq x_{n}(0)$ for $j\in\mathcal{G}_{3}$.

To prove the bounds of $\mathcal{R}_{l}$ and $\mathcal{C}_{l}$ in \eqref{eq:C_l} and \eqref{eq:R_l},  we  study the difference between $\bar{\mathbf{x}}_{[i+1]}(0)$ and $\bar{\mathbf{x}}_{[i]}(0)$ with the following three cases.
\begin{itemize}
\item[(i)]  If $i\in\mathcal{H}_{l}$, $i+1\notin\mathcal{H}_{l}$, then there must hold
\[\begin{array}{ll}
&\left\{
\begin{array}{l}
\mathcal{S}_{i}(0)=(k_{1}^{(i)}(0),0, k_{3}^{(i)}(0)), \\
\mathcal{S}_{i+1}(0) =(k_{1}^{(i)}(0)-1,1, k_{3}^{(i)}(0)) ;
\end{array}
\right.\\
\textrm{ or }&
\left\{
\begin{array}{l}
\mathcal{S}_{i}(0)=( k_{1}^{(i)}(0),1, k_{3}^{(i)}(0)), \\
\mathcal{S}_{i+1}(0)=(k_{1}^{(i)}(0),0, k_{3}^{(i)}(0)+1).
\end{array}
\right.
\end{array}\]
Taking the parameter $\mathcal{C}_{l}$ into account, we can obtain that
\begin{align}\label{lower-fluctuate-5}
\mathcal{C}_{l}=&\min_{i\in\mathcal{H}_{l},j\notin\mathcal{H}_{l}}\{|\bar{\mathbf{x}}_{[j]}(0)-\bar{\mathbf{x}}_{[i]}(0)|\}\nonumber\\
{=}&\min_{i\in\mathcal{H}_{l},i+1\notin\mathcal{H}_{l}}\{|\bar{\mathbf{x}}_{[i+1]}(0)-\bar{\mathbf{x}}_{[i]}(0)|\}\nonumber\\
{\leq}&\frac{\max\{x_{s}(0)-x_{s-1}(0),x_{s+1}(0)-x_{s}(0)\}}{m}\nonumber\\
=&\frac{\max\{D_{[s-1,s]}(0),D_{[s,s+1]}(0)\}}{m}
\end{align}
and
\begin{align*}
\mathcal{C}_{l}=&
\min_{i\in\mathcal{H}_{l},i+1\notin\mathcal{H}_{l}}\{|\bar{\mathbf{x}}_{[i+1]}(0)-\bar{\mathbf{x}}_{[i]}(0)|\}\\
\geq& \frac{\min\{D_{[s-1,s]}(0),D_{[s,s+1]}(0)\}}{m}\\
=&\frac{\beta_{s}}{m}.
\end{align*}

\item[(ii)] If $i,i+1\in\mathcal{H}_{l}$, then we have
\begin{align}\label{lower-fluctuate-4}
\max_{i,i+1\in\mathcal{H}_{l}}|\bar{\mathbf{x}}_{[i+1]}(0)-\bar{\mathbf{x}}_{[i]}(0)|
\stackrel{a)}{=}&
\max_{i,i+1\in\mathcal{H}_{l},i_{1},j_{1}\in\mathcal{G}_{k},k=1\textrm{ or }3}|\frac{x_{j_{1}}(0)-x_{i_{1}}(0)}{m}| \nonumber\\
\stackrel{b)}{\leq}&\frac{\alpha_{s}}{m}\\
\max_{i,i+1\in\mathcal{H}_{l}}|\bar{\mathbf{x}}_{[i+1]}(0)-\bar{\mathbf{x}}_{[i]}(0)|
\geq&
\min_{i,i+1\in\mathcal{H}_{l}}|\bar{\mathbf{x}}_{[i+1]}(0)-\bar{\mathbf{x}}_{[i]}(0)| \nonumber\\
\stackrel{(c)}{\geq}&\frac{1}{m}\min_{i\in\{1,2,\dots,n-1\}}D_{[i,i+1]}(0)
\end{align}
where the equality $a)$ is deduced by using the fact that there exists only one pair $(i_{1},j_{1})$ for the average $\bar{\mathbf{x}}_{[i]}(0)$ and  $\bar{\mathbf{x}}_{[i+1]}(0)$, the inequality $b)$ is obtained by using  $|x_{j_{1}}(0)-x_{i_{1}}(0)|\leq\max\{D_{[1,s-1]}(0),D_{[s+1,n]}(0)\}$ for $i_{1},j_{1}$ belonging to the same set, and
the inequality $(c)$ is obtained by using $|x_{j}(0)-x_{i}(0)|\geq\min_{i\in\mathrm{V}}D_{[i,i+1]}(0)$.

\item[(iii)] If $i,j\in\mathcal{H}_{l}$ and $k_{2}^{(i)}(0)=0$ for $\mathcal{S}_{i}(0)$, then there hold
\begin{align*}
\mathcal{R}_{l}\stackrel{a)}{\leq}&
\max_{i,j\in\mathcal{H}_{l}}\frac{1}{m}\{(m-k)D_{[s+1,n]}(0)+kD_{[1,s-1]}(0)\}\\
\leq& \alpha_{s}
\end{align*}
where we have sued $\max_{i,j\in\mathcal{G}_{1}}\{|x_{i}(0)-x_{j}(0)|\}\leq D_{[1,s-1]}(0)$ and $\max_{i,j\in\mathcal{G}_{3}}\{|x_{i}(0)-x_{j}(0)|\}\leq D_{[s+1,n]}(0)$ to obtain the inequality $a)$.
On the other hand, if $i,j\in\mathcal{H}_{l}$ and $k_{2}^{(i)}(0)=1$ for $\mathcal{S}_{i}(0)$, then
\begin{align*}
\mathcal{R}_{l}\leq&
\max_{i,j\in\mathcal{H}_{l}}\frac{1}{m}\{(m-k)D_{[s+1,n]}(0)+(k-1)D_{[1,s-1]}(0)\}\\
\leq &\frac{m-1}{m}\alpha_{s}.
\end{align*}

Similarly, the lower bound of $\mathcal{R}_{l}$ can be concluded by
\begin{align*}
\mathcal{R}_{l}\stackrel{a)}{=}&\frac{1}{m}\max_{i,j\in\mathcal{H}_{l}}|\sum_{i\in\mathcal{G}_{1}}x_{i}(0)-\sum_{j\in\mathcal{G}_{1}}x_{j}(0)|\\
\stackrel{b)}{\geq}&\frac{1}{m}\min\{D_{[1,s-1]}(0),D_{[s+1,n]}(0)\}
\end{align*}
where the equality $a)$ holds by setting $\max_{i,j\in\mathcal{H}_{l}}|\bar{\mathbf{x}}_{[i]}(0)-\bar{\mathbf{x}}_{[j]}(0)|=\frac{1}{m}|\sum_{k=1}^{m}(x_{i_{k}}(0)-x_{j_{k}}(0))|$ and the inequality $b)$ is obtained by using $|\sum_{i_{s},j_{s}\in\mathcal{G}_{1}}(x_{i_{s}}(0)-x_{j_{s}}(0))|\geq D_{[1,s-1]}(0)$, $|\sum_{i_{s},j_{s}\in\mathcal{G}_{3}}(x_{i_{s}}(0)-x_{j_{s}}(0))|\geq D_{[s+1,n]}(0)$ and $|\sum_{i_{s},j_{s}\in\mathcal{G}_{2}}(x_{i_{s}}(0)-x_{j_{s}}(0))|\geq 0$.
\end{itemize}

In summary of the above three cases, the lemma is thus concluded.  \hfill$\square$



\subsection*{D.2 Opinion Fluctuations}

In this subsection, we proceed to prove that the opinion order is preserved and node $s$  fluctuates a.s. within certain conditions based on the measure of the quotient sets.

\begin{lemma}\label{lemma-clustering-3}
Let $s\in\{n-m+1,\dots,m-1\}$. If  $x_{s}(0)\in[\underline{\alpha}_{j}(0),\overline{\alpha}_{j}(0)]$ for some $j\in\{1,2,\dots,K_{s}\}$, $\beta_{s}>3\alpha_{s}$ and the confidence bound $\eta$ satisfies $$\frac{\alpha_{s}}{m}<\eta\leq\frac{\beta_{s}}{m}-\frac{\alpha_{s}}{m-1},$$
then the order is unchanged and node $s$ fluctuates a.s..
\end{lemma}
{\it Proof.} The proof of this lemma consists of three steps.

\medskip
\noindent {\em {Step 1}.} At the first step,  we aim to show that agent $s$ has a positive probability to change its opinion values at any time $t$, and the opinion order is preserved.

Without loss of generality, we consider the case with $x_{s}(0)\in[\underline{\alpha}_{j},\overline{\alpha}_{j}]$ for some $j\in\{1,2,\dots,K_{s}\}$. At $t=0$, by Lemma \ref{lemma-clustering-1}, if $\eta>\frac{\alpha_{s}}{m}$, we then can obtain
\begin{align*}
\min_{i\in\bar{\mathrm{V}}}|\bar{\mathbf{x}}_{[i]}(0)-x_{s}(0)|\leq \min_{i\in\mathcal{H}_{j}}|\bar{\mathbf{x}}_{[i]}(0)-x_{s}(0)|\leq \frac{\alpha_{s}}{m}<\eta,\qquad \quad \forall j\in\{1,2,\ldots,C_{n}^{m}\}.
\end{align*}
This implies that agent $s$ will change its value if it selects the average opinion with the index in any $\mathcal{H}_{j}$, $j\in\{1,2,\ldots,C_{n}^{m}\}$. Thus, by Assumption 1, agent $s$ has a positive probability to change its opinion values at $t=0$.

We now show that the opinion order is preserved at $t=0$. It is noted that
\begin{align*}
&\bar{\mathbf{x}}_{[1]}(0)-x_{s-1}(0)\\
\stackrel{a)}{\geq}&\frac{m-s}{m}D_{[s-1,s+1]}(0)+\frac{1}{m}D_{[s-1,s]}(0)-\frac{s-1}{m}D_{[1,s-1]}(0)\\
\stackrel{b)}{\geq}&\frac{2m-2s+1}{m}\beta_{s}-\frac{s-1}{m}D_{[1,s-1]}(0)\\
>&\frac{\beta_{s}}{m}-\frac{\alpha_{s}}{m-1}\\
>&\eta,
\end{align*}
where the inequality $a)$ is deduced by (\ref{lower-fluctuate-1}) and $b)$ by $(m-2s+1)\beta_{s}>(m-2)\alpha_{s}>(s-1)\alpha_{s}$. Similarly, one can further obtain $x_{s+1}(0)-\bar{\mathbf{x}}_{[C_{n}^{m}]}(0)>\eta$. Thus, it follows that
$$\min_{j\neq s,\forall k\in\bar{\mathbf{V}}}|x_{j}(0)-\bar{\mathbf{x}}_{[k]}(0)|>\eta.$$
Therefore, the opinion order is preserved at $t=0$.

Next we proceed to show that agent $s$ has a positive probability to change its opinion values, and the opinion order is preserved  at $t=1$.
If $x_{s}(1)>x_{s}(0)$, $x_{s}(1)\leq x_{s}(0)+\delta\min\{\mathcal{R}_{l},\eta\}$. Similarly, all the average opinion $\bar{\mathbf{x}}_{[l]}(1)>\bar{\mathbf{x}}_{[l]}(0)$ for $k_{l}^{(2)}(0)=1$, and $\bar{\mathbf{x}}_{[l]}(1)\leq\bar{\mathbf{x}}_{[l]}(0)+\frac{\delta\min\{\mathcal{R}_{l},\eta\}}{m}$. Therefore,
\begin{align*}
&\min_{i\notin\mathcal{H}_{j}}|x_{s}(1)-\bar{\mathbf{x}}_{[i]}(1)|\\
\stackrel{a)}{>}&\min_{i\notin\mathcal{H}_{j}}|x_{s}(0)-
\bar{\mathbf{x}}_{[i]}(0)|-\frac{\delta}{m}\min\{\mathcal{R}_{l},\eta\}\\
\stackrel{b)}{>}&\eta
\end{align*}
where $a)$ is obtained by using  $x_{s}(0)\in[\underline{\alpha}_{j},\overline{\alpha}_{j}]$, $x_{s}(1)\leq x_{s}(0)+\delta\min\{\mathcal{R}_{l},\eta\}$ and $\bar{\mathbf{x}}_{[l]}(1)\leq\bar{\mathbf{x}}_{[l]}(0)+\frac{\delta\min\{\mathcal{R}_{l},\eta\}}{m}$, and the inequality $b)$ is deduced by  $\eta\leq\frac{1}{m}\min\{D_{[s-1,s]}(0),D_{[s,s+1]}(0)\}-\frac{\alpha_{s}}{m-1}$ and Lemma \ref{lemma-clustering-1}. Then we have $x_{s}(1)\in[\underline{\alpha}_{j}(1),\overline{\alpha}_{j}(1)]$ holds by Lemma \ref{lemma-clustering-1} and $x_{s}(1)<\bar{\mathbf{x}}_{[i]}(0)$ where $\bar{\mathbf{x}}_{[i]}(1)\geq\bar{\mathbf{x}}_{[i]}(0)$ if $k_{i}^{(2)}=1$, $i\in\mathcal{H}_{j}$. The similar conclusion can be obtained if $x_{s}(1)<x_{s}(0)$.
Thus, agent $s$ has a positive probability to change its value at time $1$.

Taking the opinion order at $t=1$ into consideration, we observe that
\begin{align}\label{lemma-3-1}
\min\{|\bar{\mathbf{x}}_{[1]}(0)-\mathbf{x}_{[s-1]}(0)|,|\mathbf{x}_{[s+1]}(0)-\bar{\mathbf{x}}_{[C_{n}^{m}]}(0)|\}
 -
\frac{\max_{l\in\{1,2,\dots,K_{s}\}}\{\mathcal{R}_{l}\}}{m-1}
\stackrel{a)}{\geq}\frac{\beta_{s}}{m}-\frac{\alpha_{s}}{m-1}>\eta
\end{align}
where $a)$ is deduced by using (\ref{lower-fluctuate-1}) and $\frac{\min\{\mathcal{R}_{l},\eta\}}{m-1}>\frac{\delta\eta}{m}$. This, together with Lemma \ref{lemma-clustering-1}, concludes that opinion order is preserved at time $1$.

Similarly, with the similar proof method used in Theorem \ref{allselected-converge}, we can get that the opinion order is preserved at any time. Thus, agent $s$ has a positive probability to change its opinion values at any time. Specially, our proof in $(3)$ shows that agent $s$ approaches to the upper bound of $\{\bar{\mathbf{x}}_{[i]}(t)\}$ if the selected average opinion is larger than $x_{s}(t)$, $i\in\mathcal{H}_{j}$, $t\in\mathds{N}$. While agent $s$ still have a positive probability to change its value towards the opposite direction.

\medskip
\noindent {\em {Step 2}.} At the second step, we proceed to show that that opinions of all other agents are unchanged, even if agent $s$ has a maximum range to change for $t>0$.

The extremal condition for agent $s$ to influence any other opinions is through the influence of average opinions' adjustment range, whereas we only need to judge whether the changed average opinion is in the confidence range of any other opinions. By Lemma \ref{lemma-clustering-1} and $\frac{\alpha_{s}}{m}<\eta\leq\frac{\beta_{s}}{m}-\frac{\alpha_{s}}{m-1}$, we can obtain
\begin{align}\label{lemma-3-2}
\eta\leq\frac{\beta_{s}}{m}-\frac{\alpha_{s}}{m-1}
<\min_{l\in\{1,2,\dots,K_{s}\}}\mathcal{C}_{l}-\frac{1}{m-1}\max_{l\in\{1,2,\dots,K_{s}\}}\mathcal{R}_{l}.
\end{align}

With the similar proof method to the proof of Theorem \ref{allselected-converge}, it can be seen that for any $t>0$, the maximum movable range of $x_{s}(0)$ is $\frac{m}{m-1}\mathcal{R}_{l}$, and then the maximum move range of $\bar{\mathbf{x}}_{i}(0)$ for any $i\in\bar{\mathrm{V}}$ is $\frac{1}{m-1}\mathcal{R}_{l}$.

By (\ref{lemma-3-2}), it is observed that  agent $s$ is not  influenced by average opinions out of $\mathcal{H}_{j}$. Thus, the movable range of any average opinions is not larger than $\frac{1}{m-1}\max_{l}\{\mathcal{R}_{l}\}$. With the same method of deriving the inequality (\ref{lemma-3-1}), we get that all other opinions will keep unchanged at any time.

\medskip
\noindent {\em Step 3.} At the final step, we will show that the upper limit of agent $s$ is larger than the lower limit of agent $s$ almost surely.
By Lemma \ref{lemma-clustering-1}, we have
$$
\eta>\frac{\alpha_{s}}{m}>\max_{\mathcal{S}_{i}(0)=\mathcal{S}_{i+1}(0)}|\bar{\mathbf{x}}_{[i+1]}(0)-\bar{\mathbf{x}}_{[i]}(0)|.
$$
This indicates that agent $s$ can be influenced by at least two average opinions within the index set $\mathcal{H}_{j}$. Without loss of generality, we set $\mathcal{H}_{j}=\{1,2,\dots,i_{j}\}$. In addition, agent opinion $x_{s}(0)\in[\bar{\mathbf{x}}_{[1]}(0),\bar{\mathbf{x}}_{[2]}(0))$. Assume $|\bar{\mathbf{x}}_{[i]}(0)-x_{s}(0)|\leq\eta$ for $i=1,2$ and $|\bar{\mathbf{x}}_{[j]}(0)-x_{s}(0)|>\eta$ for $j>2$. By Assumption 1, with a positive probability, the state of agent $s$ will decrease if it selects average opinion $\bar{\mathbf{x}}_{[1]}(0)$ and increases if it selects average opinion $\bar{\mathbf{x}}_{[2]}(0)$.

Towards this end, in the following we consider two cases where $k_{i}^{(2)}(0)=0$ and $k_{i}^{(2)}(0)=1$, respectively.
\begin{itemize}
\item[(i)] If $k_{i}^{(2)}(0)=0$ for $i\in\mathcal{H}_{j}$, then all average opinions $\bar{\mathbf{x}}_{[i]}(0)$ are unchanged for $i\in\mathcal{H}_{j}$. Thus, the opinion dynamics (\ref{mode}) can be rewritten as
    \begin{align*}
    x_{s}(t+1)=\left\{
    \begin{array}{ll}
    x_{s}(t)+\delta(\bar{\mathbf{x}}_{[1]}(0)-x_{s}(t)),\quad \textrm{if the selected average opinion is } [1];\\
    x_{s}(t)+\delta(\bar{\mathbf{x}}_{[2]}(0)-x_{s}(t)),  \quad \textrm{if the selected average opinion is } [2];\\
    x_{s}(t),~~~~~~~~~~~~~~~~~~~~~~~~ \quad \textrm{ otherwise. }
    \end{array}
    \right.
    \end{align*}
    This yields  that $\limsup\limits_{t\to\infty}x_{s}(t)>\liminf\limits_{t\to\infty}x_{s}(t)$ holds.
\item[(ii)] If $k_{i}^{(2)}(0)=1$ for $i\in\mathcal{H}_{j}$, then all average opinion $\bar{\mathbf{x}}_{[i]}(t)$ will change if $x_{s}(t)$ selects average opinion $\bar{\mathbf{x}}_{[1]}(t)$ for any $t\geq0$. Thus, the opinion dynamics (\ref{mode}) can be rewritten as
    \begin{align*}
    x_{s}(t)=x_{s}(0)+\delta\sum_{k=0}^{t-1}\left(1-\delta\frac{m-1}{m}\right)^{t-1-k-\sum\limits_{p=k}^{t-1}Q_{p}}\left(\bar{\mathbf{x}}_{[i]}(0)-x_{s}(0)\right)(1-Q_{k})
    \end{align*}
    where $Q_{k}=0$ if the selected average opinion index of agent $s$ belongs to $\{[1],[2]\}$, otherwise $Q_{k}=1$. This, together with the fact that $\bar{\mathbf{x}}_{[1]}(0)\leq x_{s}(0)<\bar{\mathbf{x}}_{[2]}(0)$, yields that $\limsup_{t\to\infty}x_{s}(t)>\liminf_{t\to\infty}x_{s}(t)$ holds.
\end{itemize}

\medskip
In summary of the previous analysis, the proof of \mbox{Lemma 2} is finished.
 \hfill$\square$

%

\subsection*{D.3 Fluctuation Events}

Instrumental to the subsequent analysis is the following two technical lemmas in probability theory.

\begin{lemma}\label{lemma-transfer-density} (\cite[Theorem 1.6.9]{Durrett-book} Change of variables formula)
If $\mathbf{Y}=L\mathbf{X}$ where $\mathbf{X},\mathbf{Y}\in\mathbb{R}^{1\times K}$, $L\in\mathbb{R}^{K\times K}$ is positive, and the probability densities of $\mathbf{X}$, $\mathbf{Y}$ exist, then we have
\begin{align*}
f_{Y}(y)=\frac{1}{|L|}f_{X}(L^{-1}y)
\end{align*}
where $y=(y_{1},y_{2},\dots,y_{K})'$.
\end{lemma}

\begin{lemma}\label{lemma-orderstatis} (\cite[Lemma 8.4.2]{Durrett-book})
The probability density of $X_{*}=(\mathbf{x}_{[i_{1}]}(0),\mathbf{x}_{[i_{2}]}(0),\dots,\mathbf{x}_{[i_{k}]}(0))$ satisfies
\begin{align}\label{prob-density-orderstatis}
f_{X_{*}}(\textbf{x})=\left\{
\begin{array}{ll}
\frac{n!}{(i_{1}-1)!(n-i_{k})!\prod\limits_{s=1}^{k-1}(i_{s+1}-i_{s}-1)!}x_{1}^{i_{1}-1}(1-x_{n})^{n-i_{k}}\prod\limits_{s=1}^{k-1}(x_{s+1}-x_{s})^{i_{s+1}-i_{s}-1},
\quad  \mbox{for } 0\leq x_{1}\leq \dots \leq x_{k}\leq 1;\\
0,~~~~~~~~~~~~~~~~~~~~~~~~\qquad\qquad  \mbox{otherwises}.
\end{array}
\right.
\end{align}
where $\textbf{x}=(x_{1},x_{2},\dots,x_{k})$.
\end{lemma}

With these lemmas in mind, we are now ready to prove Theorem 4.
With Lemma \ref{lemma-clustering-1} and Lemma \ref{lemma-clustering-3}, the proof of Theorem 4 reduces to find the initial condition that satisfies the condition of Lemma \ref{lemma-clustering-3}, which consists of the following three steps.

\medskip
\noindent {\em {Step 1}.} At the first step, we provide an initial condition that $x_{s}(0)\in [\underline{\alpha}_{j}(0),\overline{\alpha}_{j}(0)]$ for certain $j\in\{1,2,\dots,K_{s}\}$. We denote
\begin{align*}
&\underline{\gamma}_{1}^{k}=\frac{k-1}{m}x_{1}(0)+\frac{x_{s}(0)}{m}+\frac{m-k}{m}x_{s+1}(0),\\
&\overline{\gamma}_{1}^{k}=\frac{k-1}{m}x_{s-1}(0)+\frac{x_{s}(0)}{m}+\frac{m-k}{m}x_{n}(0),\\
&\underline{\gamma}_{0}^{k}=\frac{k}{m}x_{1}(0)+\frac{m-k}{m}x_{s+1}(0),\\
&\overline{\gamma}_{0}^{k}=\frac{k}{m}x_{s-1}(0)+\frac{m-k}{m}x_{n}(0).
\end{align*}

Given any $s\in\{n-m+1,\dots,m-1\}$ and an opinion selection tube $(k-1,1,m-k)$, we can obtain that
\begin{align}\label{lower-bound-fluctuation-1}
[\underline{\alpha}_{j}(0),\overline{\alpha}_{j}(0)]\subset[\underline{\gamma}_{1}^{k},\overline{\gamma}_{1}^{k}].
\end{align}
We then consider the opinion selection tube $(k,0,m-k)$, and can obtain that
\begin{align}\label{lower-bound-fluctuation-2}
[\underline{\alpha}_{j}(0),\overline{\alpha}_{j}(0)]\subset[\underline{\gamma}_{0}^{k},\overline{\gamma}_{0}^{k}].
\end{align}
With (\ref{lower-bound-fluctuation-1}) and (\ref{lower-bound-fluctuation-2}), it follows  the following relation of adjacent average opinion cliques:
\begin{itemize}
  \item[(i)] $\underline{\alpha}_{j+1}(0)-\overline{\alpha}_{j}(0)\geq \frac{D_{[s,s+1]}(0)-(m-1)\alpha_{s}}{m}$, if the selection tube of $\mathcal{H}_{i_{j}}$ is $(k-1,1,m-k)$ and the selection tube of $\mathcal{H}_{i_{j+1}}$ is $(k,0,m-k+1)$;
  \item[(ii)] $\underline{\alpha}_{j+1}(0)-\overline{\alpha}_{j}(0)\geq \frac{D_{[s-1,s]}(0)-(m-1)\alpha_{s}}{m}$, if the selection tube of $\mathcal{H}_{i_{j}}$ is $(k,0,m-k) $ \and the selection tube of $\mathcal{H}_{i_{j+1}}$ is $(k-1,1,m-k+1)$.
\end{itemize}
This deduces that agent $s$  fluctuates almost surely if
\begin{align}\label{lower-bound-parameters-1}
x_{s}(0)\in[\underline{\gamma}_{1}^{k},\overline{\gamma}_{1}^{k}] \textrm{ or } x_{s}(0)\in[\underline{\gamma}_{0}^{k},\overline{\gamma}_{0}^{k}]
\end{align}
which satisfies the initial conditions in Lemma \ref{lemma-clustering-3}, and thus completes the first task.

\medskip
\noindent {\em Step 2.}  At this step, we show  that agent $s$ fluctuates almost surely. The following two cases are studied, respectively.

\begin{itemize}
\item[(i)] If $s\in\mathcal{H}_{i_{1}}$, then with the similar method in Theorem \ref{allselected-converge}, we need
\begin{align}\label{lower-bound-fluctuation-3}
x_{s-1}(0)+\eta+\frac{\alpha_{s}}{m-1}<\bar{\mathbf{x}}_{[1]}(0)
\end{align}
to ensure all opinions less than $s$ can not be influenced by any average opinions. In fact, if $\eta<\frac{\beta_{s}-(m-1)\alpha_{s}}{m}-\frac{\alpha_{s}}{m-1}$, then $\eta<\frac{2m-2s+1}{m}\beta_{s}-\frac{s-1}{m}\alpha_{s}-\frac{\alpha_{s}}{m-1}$ for any $s=n-m+1,\dots,m-1$, which induces that
\begin{align*}
\eta<\frac{s-1}{m}x_{1}(0)+\frac{1}{m}x_{s}(0)+\frac{m-s}{m}x_{s+1}(0)
-x_{s-1}(0)-\frac{\alpha_{s}}{m-1}
\end{align*}
and then the inequality (\ref{lower-bound-fluctuation-3}) holds. Similarly, we can get that if $\eta<\frac{\beta_{s}-(m-1)\alpha_{s}}{m}-\frac{\alpha_{s}}{m-1}$, then $\bar{\mathbf{x}}_{[C_{n}^{m}]}(0)+\eta+\frac{\alpha_{s}}{m-1}<x_{s+1}(0)$ holds.
\item[(ii)] If $x_{s}(0)\in[\underline{\gamma}_{1}^{k},\overline{\gamma}_{1}^{k}]$, then
\begin{align}\label{lower-bound-fluctuation-4}
\frac{m-k}{m}D_{[s,s+1]}(0)<\frac{k-1}{m}D_{[1,s]}(0),\nonumber\\
\frac{k-1}{m}D_{[s-1,s]}(0)<\frac{m-k}{m}D_{[s,n]}(0).
\end{align}
Similarly, if $x_{s}(0)\in[\underline{\gamma}_{0}^{k},\overline{\gamma}_{0}^{k}]$, then
\begin{align}\label{lower-bound-fluctuation-5}
\frac{m-k}{m}D_{[s,s+1]}(0)<\frac{k}{m}D_{[1,s]}(0),\nonumber \\
\frac{k}{m}D_{[s-1,s]}(0)<\frac{m-k}{m}D_{[s,n]}(0).
\end{align}
\end{itemize}
Note that both the average opinion ranges in (\ref{lower-bound-fluctuation-1}) and (\ref{lower-bound-fluctuation-2}) are less than $\alpha_{s}$. By Lemma \ref{lemma-clustering-1}, agent $s$ can move if $\eta>\frac{\alpha_{s}}{m}$.

In summary, by Lemma \ref{lemma-clustering-3}, if
\begin{align}\label{lower-bound-parameters-2}
\frac{\alpha_{s}}{m}<\eta<\frac{\beta_{s}-(m-1)\alpha_{s}}{m}-\frac{\alpha_{s}}{m-1},
\end{align}
either (\ref{lower-bound-fluctuation-4}) holds for any $k=1,2,\dots,s$ or (\ref{lower-bound-fluctuation-5}) holds for any $k=1,2,\dots,s-1$, then we can get that opinion $s$ will fluctuate a.s..

\medskip
\noindent {\em Step 3.} Finally, we derive the lower probability bound of the initial condition $x_s(0)$. By the previous analysis, we only need to consider the parameter restrictions for the initial states under Assumptions  {\bf A1} and {\bf A2}.

By the inequalities (\ref{lower-bound-parameters-1}) and (\ref{lower-bound-parameters-2}),
\begin{align*}
\mathsf{E}_{\rm fluctuation}
\supseteq&\bigcup_{w=n-m+1}^{m-1}\Big\{\limsup x_{i}(t),\liminf x_{i}(t),i\in\mathds{V}|
\exists k\in\{1,2,\ldots,w\},\\ & x_{w}(0)\in[\underline{\gamma}_{1}^{k},\overline{\gamma}_{1}^{k}] \textrm{ or } x_{w}(0)\in[\underline{\gamma}_{0}^{k},\overline{\gamma}_{0}^{k}],\frac{\alpha_{w}}{m}<\eta<\frac{\beta_{w}-(m-1)\alpha_{w}}{m}-\frac{\alpha_{w}}{m-1}\Big\}\\
\supseteq&\Big\{\limsup x_{i}(t),\liminf x_{i}(t),i\in\mathds{V}|\exists k\in\{1,2,\ldots,s\}, \\
&\qquad x_{s}(0)\in[\underline{\gamma}_{1}^{k},\overline{\gamma}_{1}^{k}] \textrm{ or } x_{s}(0)\in[\underline{\gamma}_{0}^{k},\overline{\gamma}_{0}^{k}], \frac{\alpha_{s}}{m}<\eta<\frac{\beta_{s}-(m-1)\alpha_{s}}{m}-\frac{\alpha_{s}}{m-1}\Big\}
\end{align*}
for certain $s\in\{n-m+1,\ldots,m-1\}$. Denote opinion tube $\mathbf{X}_{s}=\left(x_{1}(0),x_{s-1}(0),x_{s}(0),x_{s+1}(0),x_{n}(0)\right)'$, where $s\in\{n-m+1,n-m+2,\dots,m-1\}$, $n<2m-1$. Then by recalling  Lemma \ref{lemma-orderstatis}, we can obtain the density function of $\mathbf{X}_{s}$ as
\begin{align}\label{lower-bound-fluctuation-8}
f_{\mathbf{X}_{s}}(x)=\left\{
\begin{array}{ll}
\frac{n!}{(s-3)!(n-s-2)!}(x_{2}-x_{1})^{s-3}(x_{5}-x_{4})^{n-s-2}, \qquad \qquad  &\mbox{for }0\leq x_{1}\leq\dots\leq x_{5}\leq 1\\
0,~~~~~~~~~~~~~~~~~~~~~~~~~~~~~~~~~~~~ &\mbox{otherwise}.
\end{array}
\right.
\end{align}

Denote
$
\mathbf{Y}_{s}=\Big(x_{1}(0),D_{[1,s-1]}(0),D_{[s-1,s]}(0),D_{[s,s+1]}(0),D_{[s+1,n]}(0)\Big)'.
$
Then $\mathbf{Y}_{s}=L_{s}\mathbf{X}_{s}$ where
\begin{align*}
L_{s}=\left(
\begin{array}{ccccc}
1 & 0 & 0 & 0 & 0\\
-1 & 1 & 0 & 0 & 0\\
0 & -1 & 1 & 0 & 0\\
0 & 0 & -1 & 1 & 0\\
0 & 0 & 0 & -1 & 1
\end{array}
\right).
\end{align*}
Further by Lemma \ref{lemma-transfer-density}, we can obtain that $f_{\mathbf{Y}_{s}}(\mathbf{y})=\frac{1}{|L_{s}|}f_{\mathbf{X}_{s}}(L_{s}^{-1}\mathbf{y})=f_{\mathbf{X}_{s}}(L_{s}^{-1}\mathbf{y})$ where $\mathbf{y}=(y_{1},y_{2},y_{3},y_{4},y_{5})$, $0\leq y_{i}< 1, \sum_{i=1}^{5}y_{i}\leq 1$. Thus, the density function (\ref{lower-bound-fluctuation-8}) can be transfered into
\begin{align*}
f_{\mathbf{Y}_{s}}(\mathbf{y})=\left\{
\begin{array}{ll}
\frac{n!}{(s-3)!(n-s-2)!}y_{2}^{s-3}y_{5}^{n-s-2},~~~  \qquad\qquad& \mbox{for } 0\leq x_{1}\leq\dots\leq x_{5}\leq 1;\\
0,~~~~~~~~~~~~~~~~~~~~~~~~~~~~\qquad  &\mbox{otherwise}.
\end{array}
\right.
\end{align*}

Therefore, $\mathbb{P}\{\mathsf{E}_{\rm fluctuation}\}\geq\int_{\mathpzc{B}_{1}\cup\mathpzc{B}_{2}}f_{\mathbf{Y}_{s}}(\mathbf{y})d\mathbf{y}$ where
\begin{align*}
\mathpzc{B}_{1}=\Big\{&\max\{y_{2},y_{5}\}<m\eta,
\min\{y_{3}, y_{4}\}>\frac{m^2-m+1}{m-1}\max\{y_{2},y_{5}\}+m\eta,\\
&\frac{m-k}{m}y_{4}<\frac{k-1}{m}(y_{2}+y_{3}),
\frac{k-1}{m}y_{3}<\frac{m-k}{m}(y_{4}+y_{5}) \Big\}
\end{align*}
and
\begin{align*}
\mathpzc{B}_{2}= \Big\{&\max\{y_{2},y_{5}\}<m\eta,
\min\{y_{3},y_{4}\}>\frac{m^2-m+1}{m-1}\max\{y_{2},y_{5} \}+m\eta,\\
&\frac{m-k}{m}y_{4}<\frac{k}{m}(y_{2}+y_{3}),
\frac{k}{m}y_{3}<\frac{m-k}{m}(y_{4}+y_{5})\Big\}\,.
\end{align*}

We can verify from Lemma \ref{lemma-orderstatis} that
\begin{align*}
\int_{\mathpzc{B}_{1}\cup\mathpzc{B}_{2}}f_{\mathbf{Y}_{s}}(\mathbf{y})d\mathbf{y}
\geq&\frac{2n!}{(s-3)!(n-s-2)!}\int_{0}^{1-2m\eta-\frac{2m^{3}\eta}{m-1}}dy_{1}\int_{0}^{m\eta}y_{2}^{s-3}dy_{2}\\
&\int_{y_{2}}^{m\eta}y_{5}^{n-s-2}dy_{5}
\int_{\frac{m^{2}-m+1}{m-1}y_{5}+m\eta}^{1-2m\eta-\frac{m^{3}\eta}{m-1}}dy_{4}
\int_{\frac{m-k}{k-1}y_{4}-m\eta}^{\frac{m-k}{k-1}(y_{4}+m\eta)}dy_{3}\\
>&0
\end{align*}
and
\begin{align*}
&\mathbb{P}\{\mathpzc{B}_{s}^{(1)}\cup\mathpzc{B}_{s}^{(1)}\cup\mathpzc{B}_{s}^{(u_{l})}\}\\
&\geq\frac{2n!(1-2m\eta-\frac{2m^{3}\eta}{m-1})(1-3m\eta-\frac{m^{3}\eta}{m-1})\frac{m}{k}(m\eta)^{n-2}}{(s-3)!(n-s-2)!}
 \left(\frac{1}{(s-2)(n-3)}-\frac{m\eta}{(s-2)(n-2)}\right) \\
&>0.
\end{align*}
We thus have
\begin{align*}
\mathbb{P}\{\mathsf{E}_{\rm fluctuation}\}\geq\int_{\mathpzc{B}_{1}\cup\mathpzc{B}_{2}}f_{\mathbf{Y}_{s}}(\mathbf{y})d\mathbf{y}>0.
\end{align*}
Particularly, it is noted that the above inequality requires the confidence bound $\eta$ to satisfy
\begin{align*}
\eta<\min\{\frac{1}{2m+\frac{2m^{3}}{m-1}},\frac{1}{3m+\frac{m^{3}}{m-1}}\}=\frac{1}{2m+\frac{2m^{3}}{m-1}}
\end{align*}
with $m\geq 4$. The proof is completed.
 \hfill$\square$
\subsection*{E. Proof of Theorem 5}

Fix the agent indexes by letting $x_{i}(0)\triangleq \mathbf{x}_{[i]}(0)$ for $i\in\mathrm{V}$ and let $s\in\{n-m+1,\dots,m-1\}$. We denote $\mathcal{G}_{1}(0)=\{[1],\dots,[K-1]\}$, $\mathcal{G}_{2}(0)=\{[K]\}$ and $\mathcal{G}_{3}(0)=\{[K+1],\dots,[n]\}$ where $n-m+2\leq K\leq m-1$, and set $\mathcal{S}_{i}(0)=(k_{1}^{(i)}(0),k_{2}^{(i)}(0),k_{3}^{(i)}(0))$ as the same definition in Appendix D for $i\in\bar{\mathrm{V}}$.

With these preliminaries, we now proceed to prove this theorem, consisting of the following steps.

\medskip
\noindent {\em Step 1.} At this step, we aim to prove that $\min_{|\bar{\mathbf{x}}_{[k]}(0)-\bar{\mathbf{x}}_{[j]}(0)|>0}\{|\bar{\mathbf{x}}_{[k]}(0)-\bar{\mathbf{x}}_{[j]}(0)|\}=\frac{2\beta}{m}(m-K+2)$, and $x_{k}(0)$ keeps unchanged at time $0$ for $1\leq k\leq K-1$, $K+1\leq k\leq n$.

For any $x_{K}(0)\in[\frac{1}{2}-\beta,\frac{1}{2}+\beta]$, we have by (\ref{lower-fluctuate-1})
\begin{align*}
&\min \bar{\mathbf{x}}_{[1]}(0)=\frac{m-K+1}{m}(\frac{1}{2}-\beta), \\
&\max \bar{\mathbf{x}}_{[1]}(0)=\frac{m-K+1}{m}(\frac{1}{2}+\beta),\\
&\min \bar{\mathbf{x}}_{[2]}(0)=\frac{m-K+2}{m}(\frac{1}{2}-\beta), \\
&\max \bar{\mathbf{x}}_{[2]}(0)=\frac{m-K+2}{m}(\frac{1}{2}+\beta)\\
& \ldots \\
&\min \bar{\mathbf{x}}_{[z]}(0)=1-\frac{m-K+1}{m}(\frac{1}{2}+\beta), \\
&\max \bar{\mathbf{x}}_{[z]}(0)=1-\frac{m-K+1}{m}(\frac{1}{2}-\beta).
\end{align*}
Then, by calculating $\min \bar{\mathbf{x}}_{[k+1]}(0)-\max \bar{\mathbf{x}}_{[k]}(0)$ and $m\leq\frac{2}{3}n$, we obtain
\begin{small}
\begin{align*}
&\min_{|\bar{\mathbf{x}}_{[k]}(0)-\bar{\mathbf{x}}_{[j]}(0)|>0}\{|\bar{\mathbf{x}}_{[k]}(0)-\bar{\mathbf{x}}_{[j]}(0)|\}\\
=&\min\{\frac{1-2\beta(2m-2(K-1)+3)}{2m},\frac{2\beta}{m}(m-K+2)\}\\
=&\frac{2\beta}{m}(m-K+2)<\eta.
\end{align*}
\end{small}
With this in mind, we further consider the following two cases.
\begin{itemize}
  \item[(i)] If $m$ is even, then there must exist an index $[j]$ such that $\mathcal{N}_{[j]}(0)$ includes $\frac{m}{2}$ selections from the index set $\mathcal{G}_{1}(0)$ and $\frac{m}{2}$ selections from $\mathcal{G}_{3}(0)$.
  \item[(ii)] If $m$ is odd, then we can get that there exists an index $[k]$ such that $\mathcal{N}_{[k]}(0)$ includes $\frac{m-1}{2}$ selections from the index set $\mathcal{G}_{1}(0)$, $\frac{m-1}{2}$ selections from $\mathcal{G}_{3}(0)$ and one from $\mathcal{G}_{2}(0)$.
\end{itemize}

In light of the above both cases, it can be concluded that there always exists at least one index such that $\bar{\mathbf{x}}_{[K]}(0)=\frac{1}{2}$. Besides, if $\mathcal{S}_{i}(0)=\mathcal{S}_{j}(0)$, then $\bar{\mathbf{x}}_{[i]}(0)=\bar{\mathbf{x}}_{[j]}(0)$ because all opinion values in the same index group are the same. By the initial values setting, we have
\begin{align*}
\bar{\mathbf{x}}_{[1]}(0)\in(0,\frac{1}{2}),\quad \bar{\mathbf{x}}_{[z]}(0)\in(\frac{1}{2},1).
\end{align*}
Besides, it follows that
\begin{align}\label{special-initial-1}
\min_{k\neq K,j\in\bar{\mathrm{V}}}\{|\mathbf{x}_{[k]}(0)-\bar{\mathbf{x}}_{[j]}(0)|\}=\frac{m-K+1}{m}(\frac{1}{2}-\beta)>\eta,
\end{align}
which implies that $\mathbf{x}_{[i]}(0)$ keeps unchanged for $i\neq K$.

\medskip
\noindent {\em {Step 2}.}
By (\ref{special-initial-1}), the opinion order is preserved at time $0$, and $x_{s}(1)=x_{s}(0)$ for $s\neq K$. Besides, there holds
\begin{align*}
\max\{x_{K}(1)-x_{K}(0)\}\leq\delta\max\{|\bar{\mathbf{x}}_{[j]}(0)-x_{K}(0)|\}\leq\delta\eta.
\end{align*}
Similarly, at $t=1$, it can be deduced that
\begin{align*}
\min_{k\neq K,j\in\bar{\mathrm{V}}}\{|x_{k}(1)-\bar{\mathbf{x}}_{[j]}(1)|\}&\geq\frac{m-K+1}{m}(\frac{1}{2}-\beta)-\frac{\delta\eta}{m}\\
&>\eta,
\end{align*}
Thus, $x_{i}(2)=x_{i}(1)$ holds for $i\neq K$. Furthermore, it is observed that
\begin{align*}
&|\mathbf{x}_{[K]}(2)-x_{K}(1)|\\
\leq&\delta\max_{|\bar{\mathbf{x}}_{[j]}(1)-x_{K}(1)|\leq\eta}
\left\{|\bar{\mathbf{x}}_{[j]}(0)+\frac{\delta\eta}{m}-x_{K}(0)-\delta(\bar{\mathbf{x}}_{[k]}
(0)-x_{K}(0))|\right\}\\
\leq&\delta\left(1-\delta+\frac{\delta}{n}\right)\eta
\end{align*}
where $\bar{\mathbf{x}}_{[j]}(0)=\bar{\mathbf{x}}_{[k]}(0)$. Therefore, $x_{s}(1)$ keeps unchanged for $s\neq K$. Similarly, we can get that $x_{i}(2)$ keeps unchanged for $i\neq K$ because
\begin{align*}
\min_{k\neq K,j\in\bar{\mathrm{V}}}\{|x_{k}(2)-\bar{\mathbf{x}}_{[j]}(2)|\}
\geq\frac{m-K+1}{m}(\frac{1}{2}-\beta)-\frac{\delta\eta}{m}-
\frac{\delta}{m}\left(1-\delta+\frac{\delta}{n}\right)\eta>\eta.
\end{align*}

\noindent {\em Step 3.} At this step, we proceed to show that $x_{i}(t)$, $i\neq K$ keeps unchanged for any $t\in\{0,1,2,\dots\}$.

Recursively, we assume that $x_{i}(t-1)$ keeps unchanged at time $t-1$ for $i\neq K$ and $|x_{K}(t)-x_{K}(t-1)|\leq\delta(1-\delta+\frac{\delta}{m})^{t-1}\eta$. With $\beta<\eta$ and $\eta<\frac{1}{6+\frac{4n}{m(n-1)}}$, one can see that $x_{i}(t)$ keeps unchanged at time $t-1$ for $i\neq K$ because
\begin{align*}
\min_{k\neq K,j\in\bar{\mathrm{V}}}\{|x_{k}(t-1)-\bar{\mathbf{x}}_{[j]}(t-1)|\}
\geq\frac{m-K+1}{m}(\frac{1}{2}-\beta)-
\frac{\delta\eta}{m}\sum_{i=0}^{t-2}(1-\delta+\frac{\delta}{n})^{i}>\eta.
\end{align*}

Along this way, as $t\to\infty$,  one can get that $\eta<\frac{1}{2}-\frac{n+mn-m}{(m-K+1)(n-1)}\eta$ and  $\frac{m-K+1}{m}(\frac{1}{2}-\beta)-\frac{n}{m(n-1)}\eta>\eta$.
As a result, there holds
$$
\sup_{t\geq 1}\{|x_{K}(t)-x_{K}(0)|\}=\sum_{k=0}^{\infty}\delta\eta(1-\delta+\frac{\delta}{n})^{k}=\frac{n\eta}{n-1}$$
which implies $x_{i}(t)=x_{i}(0)$ if $i\neq K$ where $\min_{k\neq K,j\in\bar{\mathrm{V}}}\{|x_{k}(t)-\bar{\mathbf{x}}_{[j]}(t)|\}>\eta$ , $t\in\mathds{N}$.

\medskip
\noindent {\em {Step 4}.} Finally, we prove that $\limsup\limits_{t\to\infty}x_{K}(t)>\liminf\limits_{t\to\infty}x_{K}(t)$.

By the definition of $\mathcal{S}_{i}(0)$, $i\in\bar{\mathrm{V}}$, there exists a $i_{0}\in\bar{\mathrm{V}}$ such that $0\leq x_{K}(0)-\bar{\mathbf{x}}_{[i_{0}]}(0)<\eta$ and another $j_{0}\in\bar{\mathrm{V}}$ such that $0\leq\bar{\mathbf{x}}_{[j_{0}]}(0)-x_{K}(0)<\eta$. In fact, we can set
\begin{align*}
\mathcal{S}_{i_{0}}(0)=\left\{
\begin{array}{ll}
(\frac{m+1}{2},0,\frac{m-1}{2}), \qquad \textrm{ if } m \textrm{ is odd;}\\
(\frac{m}{2},1,\frac{m}{2}-1), \qquad\,\,\textrm{ if } m \textrm{ is even},
\end{array}
\right.
\end{align*}
and
\begin{align*}
\mathcal{S}_{j_{0}}(0)=\left\{
\begin{array}{ll}
(\frac{m-1}{2},0,\frac{m+1}{2}), \qquad\textrm{ if } m \textrm{ is odd;}\\
(\frac{m}{2}-1,1,\frac{m}{2}), \qquad\,\, \textrm{ if } m \textrm{ is even}.
\end{array}
\right.
\end{align*}

By the order preservation of $\{\bar{\mathbf{x}}_{[j]}(t)\}$ for $j\in K$ at {\em Step 3}, we can obtain that $\max_{j\in\bar{\mathrm{V}}}\{|\bar{\mathbf{x}}_{[j]}(t)-\bar{\mathbf{x}}_{[j]}(0)|\}\leq\frac{\eta}{m-1}$. At time $t=1$, we will show that $x_{K}(1)$ will increase or decrease with a positive probability. In fact, there always exists ${k_{0}}\in\bar{\mathrm{V}}$ such that
\begin{align*}
\mathcal{S}_{{k_{0}}}(0)=\left\{
\begin{array}{ll}
(\frac{m-1}{2},1,\frac{m-1}{2}), \qquad\textrm{ if } m \textrm{ is odd;}\\
(\frac{m}{2},0,\frac{m}{2}), \qquad\qquad  \textrm{ if } m \textrm{ is even}.
\end{array}
\right.
\end{align*}
Thus, if $m$ is odd,
\begin{align*}
x_{K}(1)=\left\{
\begin{array}{ll}
(1-\delta)x_{K}(0)+\delta(\frac{1}{2}-\frac{1}{2m}),  \qquad\qquad\quad &\textrm{if the selected average is }\bar{\mathbf{x}}_{[i_{0}]}(0)\\
(1-\delta)x_{K}(0)+\delta(\frac{1}{2}+\frac{1}{2m}),  \qquad\qquad\quad &\textrm{if the selected average is }\bar{\mathbf{x}}_{[j_{0}]}(0)\\
(1-\delta+\frac{\delta}{m})x_{K}(0)+\delta(\frac{1}{2}-\frac{1}{2m}),  \qquad\qquad\quad &\textrm{if the selected average is }\bar{\mathbf{x}}_{[k_{0}]}(0),
\end{array}
\right.
\end{align*}
and then we analyze certain limits in the following six cases:

\begin{itemize}
  \item[(i)] If $m$ is an odd number and the selected tube is always $S_{i_{0}}(t)$, $t\geq 0$, then
\begin{align*}
x_{K}(t)=&(1-\delta)^{t}x_{K}(0)+\delta(\frac{1}{2}-\frac{1}{2m})\sum\limits_{\tau=0}^{t-1}(1-\delta)^{t-1}
\to \frac{1}{2}-\frac{1}{2m} \textrm { as } t \to\infty;
\end{align*}
  \item[(ii)] If $m$ is an even number and the selected tube is always $S_{i_{0}}(t)$, $t\geq 0$, then
\begin{align*}
x_{K}(t)=&(1-\delta+\frac{\delta}{m})^{t}x_{K}(0)+\delta(\frac{1}{2}-\frac{1}{m})\sum\limits_{\tau=0}^{t-1}(1-\delta+\frac{\delta}{m})^{\tau} \to \frac{m}{m-1}(\frac{1}{2}-\frac{1}{m}) \textrm { as } t \to\infty;
\end{align*}
  \item[(iii)] If $m$ is an odd number and the selected tube is always $S_{j_{0}}(t)$, $t\geq 0$, then
\begin{align*}
\mathbf{x}_{[K]}(t)=&(1-\delta)^{t}\mathbf{x}_{[K]}(0)+\delta(\frac{1}{2}+\frac{1}{2m})\sum\limits_{\tau=0}^{t-1}(1-\delta)^{t-1}\to \frac{1}{2}+\frac{1}{2m} \textrm { as } t \to\infty;
\end{align*}
  \item[(iv)] If $m$ is an even number and the selected tube is always $S_{j_{0}}(t)$, $t\geq 0$, then
\begin{align*}
x_{K}(t)=&(1-\delta+\frac{\delta}{m})^{t}x_{K}(0)+\frac{\delta}{2}\sum\limits_{\tau=0}^{t-1}(1-\delta+\frac{\delta}{m})^{\tau}
\to \frac{m}{2(m-1)} \textrm { as } t \to\infty;
\end{align*}
  \item[(v)] If $m$ is an odd number and the selected tube is always $S_{k_{0}}(t)$, $t\geq 0$, then
\begin{align*}
x_{K}(t)=&(1-\delta+\frac{\delta}{m})^{t}x_{K}(0) +\frac{\delta(m-1)}{2m}\sum\limits_{\tau=0}^{t-1}(1-\delta+\frac{\delta}{m})^{\tau}\to \frac{1}{2} \textrm { as } t \to\infty;
\end{align*}

  \item[(vi)] If $m$ is an even number and the selected tube is always $S_{k_{0}}(t)$, $t\geq 0$, then
\begin{align*}
x_{K}(t)=&(1-\delta)^{t}x_{K}(0)+\frac{\delta}{2}\sum\limits_{\tau=0}^{t-1}(1-\delta)^{t-1} \to \frac{1}{2} \textrm { as } t \to\infty.
\end{align*}
\end{itemize}

Note that if the selected average value of $x_{K}(t)$ is $\bar{\mathbf{x}}_{[i_{0}]}(t)$, $t=0,1,\dots,T$, then
\begin{align*}
\bar{\mathbf{x}}_{[i_{0}]}(t)=\left\{
\begin{array}{ll}
\frac{1}{2}-\frac{1}{2m},~~~~~~~~~~~~~~~~~~ &\mbox{if $m$ is odd};\\
\frac{1}{2}-\frac{1}{m}+\frac{1}{m}x_{K}(t),~~~~~~ &\mbox{if $m$ is even}.
\end{array}
\right.
\end{align*}
Besides, if the selected average value of $x_{K}(t)$ is $\bar{\mathbf{x}}_{[k_{0}]}(t)$, $t=0,1,\dots,T$, then
\begin{align*}
\bar{\mathbf{x}}_{[k_{0}]}(t)=\left\{
\begin{array}{ll}
\frac{1}{2}-\frac{1}{2m}+\frac{1}{m}x_{K}(t),~~~~ &\mbox{if $m$ is odd};\\
\frac{1}{2},~~~~~~~~~~~~~~~~~~~~~~~~~~~ &\mbox{if $m$ is even}.
\end{array}
\right.
\end{align*}
and if the selected average value of $x_{K}(t)$ is $\bar{\mathbf{x}}_{[j_{0}]}(t)$, $t=0,1,\dots,T$, then
\begin{align*}
\bar{\mathbf{x}}_{[j_{0}]}(t)=\left\{
\begin{array}{ll}
\frac{1}{2}+\frac{1}{2m},~~~~~~~~~~~~ \mbox{if $m$ is odd};\\
\frac{1}{2}+\frac{1}{m}x_{K}(t),~~~~~~ \mbox{if $m$ is even}.
\end{array}
\right.
\end{align*}

Now we proceed to study  whether the opinion changes values. With $\beta<\eta$,  we have
\[
    |x_{K}(t)-\bar{\mathbf{x}}_{[i_{0}]}(t)|=
\left\{
\begin{array}{ll}
|(\frac{1}{2}-\frac{1}{2m})(1-\delta)-x_{K}(0)|(1-\delta)^{t}, \qquad\qquad\qquad\,\,\,\, &\mbox{if $m$ is odd};\\
|(\frac{1}{2}-\frac{1}{m})(1-\delta+\frac{\delta}{m})-\frac{m-1}{m}x_{K}(0)|(1-\delta+\frac{\delta}{m})^{t},\quad\qquad  &\mbox{if $m$ is even},
\end{array}
\right.
\]
\[
|x_{K}(t)-\bar{\mathbf{x}}_{[j_{0}]}(t)|=
\left\{
\begin{array}{ll}
|(\frac{1}{2}+\frac{1}{2m})(1-\delta)-x_{K}(0)|(1-\delta)^{t}, \qquad\qquad\qquad\,\,\,\, &\mbox{if $m$ is odd};\\
|\frac{1}{2}(1-\delta+\frac{\delta}{m})-\frac{m-1}{m}x_{K}(0)| (1-\delta+\frac{\delta}{m})^{t},\qquad\quad &\mbox{if $m$ is even},
\end{array}
\right.
\]
and
\[
|x_{K}(t)-\bar{\mathbf{x}}_{[k_{0}]}(t)|=
\left\{
\begin{array}{ll}
|\frac{m-1}{2m}(1-\delta+\frac{\delta}{m})-\frac{m-1}{m}x_{K}(0)|(1-\delta+\frac{\delta}{m})^{t}\,, \quad\quad &\mbox{if $m$ is odd};\\
|\frac{1}{2}-x_{K}(0)|(1-\delta)^{t}, \quad &\mbox{if $m$ is even}.
\end{array}
\right.
\]

It is clear that $|x_{K}(t)-\bar{\mathbf{x}}_{[i_{0}]}(t)|<\eta$, $|x_{K}(t)-\bar{\mathbf{x}}_{[j_{0}]}(t)|<\eta$ and $|x_{K}(t)-\bar{\mathbf{x}}_{[k_{0}]}(t)|<\eta$. Moreover,
by combining the above three equations  together, it is concluded that $x_{K}(t)$ can always change values when it selects $i_0$, $j_0$ and $k_0$.
Therefore, for any mixed selection of $\{\bar{\mathbf{x}}_{[s]}(t),s\leq k_{0}\}$, we get that $x_{K}(t)\in[\bar{\mathbf{x}}_{[i_{0}]}(t),\bar{\mathbf{x}}_{[k_{0}]}(t)]\in(\frac{m}{m-1}(\frac{1}{2}-\frac{1}{m}),\frac{1}{2})$. With a similar method, for any mixed selection of $\{\bar{\mathbf{x}}_{[s]}(t),s\geq k_{0}\}$, we can get that $\max\{\bar{\mathbf{x}}_{[j_{0}]}(t)-\bar{\mathbf{x}}_{[k_{0}]}(t)\}<\frac{1}{2m}(1+\frac{1}{m})<\eta$ for the extremal condition that $\bar{\mathbf{x}}_{[j_{0}]}(t)$ is the selected average of agent $K$. Also, $x_{K}(t)\in[\bar{\mathbf{x}}_{[k_{0}]}(t),\bar{\mathbf{x}}_{[j_{0}]}(t)]\in(\frac{1}{2},\frac{1}{2}(1+\frac{1}{m-1}))$.

Based on above analysis, the lower limit is not larger than any given limits and the upper limit is not smaller than any given ones, we get that
\begin{align*}
&\limsup_{t\to\infty}x_{K}(t)\geq\frac{1}{2}(1+\frac{1}{m})\\
&\liminf_{t\to\infty}x_{K}(t)\leq\frac{1}{2}(1-\frac{1}{m}).
\end{align*}
The proof is thus completed.  \hfill$\square$

\end{document}